\documentclass[a4paper,12pt]{article}
\usepackage{amsmath}
\usepackage{amsfonts}
\usepackage{mathtools}
\usepackage{authblk}
\usepackage{graphicx,epsfig}
\usepackage{float}
\usepackage{caption}
\usepackage{subcaption}
\usepackage{amssymb}
\usepackage{color}
\usepackage{physics}
\usepackage{cite}
\usepackage{mathtools}
\usepackage{comment}
\usepackage{geometry}
\usepackage{array}
\usepackage{tabulary}
\usepackage{wrapfig}
\usepackage[normalem]{ulem}
\numberwithin{equation}{section}
\title{Time delay as a probe of multiple photon spheres}
\author[1,2]{Kajol Paithankar\footnote{kajol.paithankar@iiap.res.in}\,}
\author[1,2]{Sanved Kolekar\footnote{sanved.kolekar@iiap.res.in}\,}
\affil[1]{Indian Institute of Astrophysics, Koramangala II Block, Bangalore 560034, India}
\affil[2]{Pondicherry University, R.V. Nagar, Kalapet 605014, Puducherry, India}

\begin{document}
\maketitle

\begin{abstract}

Black hole shadow images are primarily determined by the properties of photon spheres and can exhibit degeneracies across different spherically symmetric spacetime geometries. We show that time delay observables associated with higher-order images of transient sources provide a robust probe to break such degeneracies in spacetimes admitting multiple photon spheres. Adopting a theory-agnostic, parametrized, static, spherically symmetric framework that captures the generic features of double-peak effective potentials, we investigate photon geodesics and quantify them in terms of angular deflection, travel time, and the order of the image. We identify distinctive signatures of trajectories probing the region between the unstable photon spheres. In particular, we find that these trajectories are characterized by the nontrivial temporal behavior, including a minimum travel time, a minimum angular deflection, and a characteristic triplet structure of higher-order images with a specific arrival sequence. We further show that the influence of the depth of the potential well, between the two photon spheres, on the observed time delays provides a direct handle on otherwise inaccessible regions of the spacetime. Although we consider a parametrized framework, these identified signatures are primarily influenced by the potential structure and will therefore occur in any spacetime with a similar double-peak effective potential. Our results highlight that time-domain lensing observables encode information beyond static shadow images and offer a promising avenue for probing the structure of compact objects and the strong-field regime of gravity.
    
\end{abstract}

\section{Introduction}

Recent horizon-scale observations of the supermassive black holes M87* and Sgr A* by the Event Horizon Telescope (EHT) have established the black hole shadows as a powerful probe of strong-field gravity \cite{EHT BH image M87, EHT BH image Milkyway}. These observational advances motivated systematic efforts to test general relativity, competing theories, and environmental effects in the near-horizon regime. At the same time, they have reinforced important limitations, particularly the degeneracies that arise when a single characteristic scale provided by the shadow diameter is used to infer the underlying spacetime geometry. Future instruments, including next-generation EHT (ngEHT) and missions such as the Black Hole Explorer (BHEX), are expected to achieve significantly improved resolution, enabling the detection of fine structures such as photon sub-rings and their temporal behavior \cite{ngEHT1, ngEHT2, ngEHT3, BHEX1, BHEX2}. In this context, it is crucial to examine how black hole images may differ under varying astrophysical conditions, and to identify observables that offer more sensitive and robust probes of the underlying geometry.

Central to these observations is the concept of the photon sphere, which is a surface of unstable circular null geodesics that defines the edge of the black hole shadow. In the standard paradigm of asymptotically flat, spherically symmetric black holes, such as the Schwarzschild and Reissner-Nordström (RN) solutions, the effective potential encountered by photons typically exhibits a single local maximum outside the event horizon. This single peak governs the light deflection angle, which diverges logarithmically as the impact parameter approaches the critical value, generating a sequence of self-similar relativistic images \cite{Sch lensing Virbhadra, logarithmic divergence Bozza, Time delay as distance estimator}. However, recent explorations of non-vacuum solutions in general relativity (GR) and modified gravity theories have revealed spacetimes possessing double-peak effective potentials, allowing the existence of multiple photon spheres. Such potentials have been identified in exact and numerical spacetimes including traversable wormholes \cite{Traversable wormhole solution}, hairy Schwarzschild black holes \cite{Hairy Sch solution}, hairy Reissner-Nordström black holes \cite{Scalarized RN BH Solution}, black holes surrounded by dark matter halo \cite{DM halo}, dyonic black holes \cite{dyonic black holes} as well as in parametrized geometries such as Johannsen-Psaltis (JP) metric \cite{JP metric}. The physical viability of various black hole solutions with multiple photon spheres has been assessed using the energy conditions, and it has been found that such solutions are not exotic \cite{BH with multiple PS}.

The presence of a double-peak potential leads to nontrivial modifications of the geodesic dynamics. Geometrically, the presence of two unstable photon spheres implies the existence of a stable photon sphere, called the anti-photon sphere \cite{anti-photon terminology}. The anti-photon sphere is located at the minimum between the two potential peaks. While photons trapped at the anti-photon sphere are generally not visible to distant observers, the complete structure of the potential generates distinctive optical signatures, such as multiring structures and triplets of higher-order relativistic images, which differ significantly from the standard infinite sequence observed in Schwarzschild spacetime \cite{hairy Schwarzschild, Hairy RN lensing, DM halo shadow, Lensing by multiple photon sphere, multiring structure, Traversable wormholes, thin-shell wormholes}. In the context of interferometric observations, the interference of light rays orbiting the two unstable photon spheres has been shown to produce characteristic beat signals in the visibility amplitude, breaking the universal cascade of damped oscillations associated with single-photon sphere models \cite{beat pattern Sch, beat pattern RN}. Furthermore, the presence of the potential well between the peaks can lead to long-lived quasinormal modes and the gravitational wave echoes \cite{QNM Echoes of Sch Hernquist BHs, QNMs of BH encircled by a gravitating thin disk, QNM Echoes of hairy BHs, QNM Echoes of dyonic BHs, QNM Echoes of hairy RN}.

While numerical ray tracing has been extensively used to simulate the visual appearance of such objects, a critical aspect of the time delay of images in these spacetimes remains less explored. Time delay studies, historically established through the Shapiro effect as a fundamental test of gravity \cite{Shapiro}, are now pivotal in the strong-field regime. The time-resolved observations from the ngEHT and BHEX missions will allow us to study transient phenomena, such as light echoes, i.e., the higher-order images that arrive with a specific time delay relative to the direct image. In strong gravitational lensing, time-domain observations provide a complementary and crucial probe of the spacetime metric. A method to infer light echoes and echo time delays through interferometric observations in the background of Kerr geometry has been developed in \cite{Light echoes}. In the case of Kerr-Newman black holes, it has been found that the charge of the black hole enhances the time delay between prograde and retrograde orbits in the equatorial plane \cite{Time Delay Kerr Newman}.

In this paper, we explore the effects of multiple photon spheres on the strong gravitational lensing and the corresponding time delays within a parametrized spacetime. In section \ref{Spacetime model}, we introduce the background spacetime and discuss the features of the double-peak effective potential encountered by null geodesics in section \ref{Null geodesics}. Such an effective potential can produce a black hole image with two photon rings, whose sizes depend on the magnitudes of the potential at the two peaks. Our goal is to explore the time delay between relativistic images and indicate that it can serve as a key observable to break the degeneracy among the intrinsic geometric parameters. We set up the problem of parameter degeneracy in section \ref{double peak potentials}, where we choose two sets of black hole parameters that generate identical photon rings in the images of the black hole shadows. In these background spacetimes, we investigate how the deflection angle and the arrival times of relativistic images are influenced by the double-peak potential and compare it with the Schwarzschild case in section \ref{comparison with SCh}. In section \ref{Time delay of BHs}, we consider a fixed point source and examine in detail the photon trajectories and time delays caused by the double-peak potential in the background of the two black hole spacetimes. Further, to isolate the effect of the depth of the potential well, we investigate the photon trajectories in a background of an exotic compact object in section \ref{ECO}. The conclusions are discussed in section \ref{discussion section}. By analyzing the time-delay signatures, we provide a distinguishing feature of the imprint of double-peak potentials.

The metric signature adopted is $(-,+,+,+)$ and natural units $G=\hbar=c=1$ are used.

\section{The Setup}\label{Spacetime model}

The primary aim of this work is to investigate time variability, in particular the time delays, arising from the presence of double-peak effective potentials in static, spherically symmetric, asymptotically flat spacetimes. While one may analyze the specific solutions discussed in the previous section, we instead adopt a parameterized metric. This choice of metric allows us to work in a convenient framework that incorporates the relevant features of the effective potential. Although the numerical analysis is performed within this parameterized background, the resulting time-delay signatures are not limited to it. The null geodesic motion is dominantly affected by the potential encountered by photons, and its effect will be imprinted in the prominent features of the observables. Consequently, the qualitative features of the time delay identified in this work will extend to any black hole spacetime sharing the same underlying potential structure, regardless of the specific choice of model or the theory of gravity. In this sense, our results are model-independent.
 
\begin{figure}[h!]
    \centering
    \includegraphics[width=0.5\linewidth]{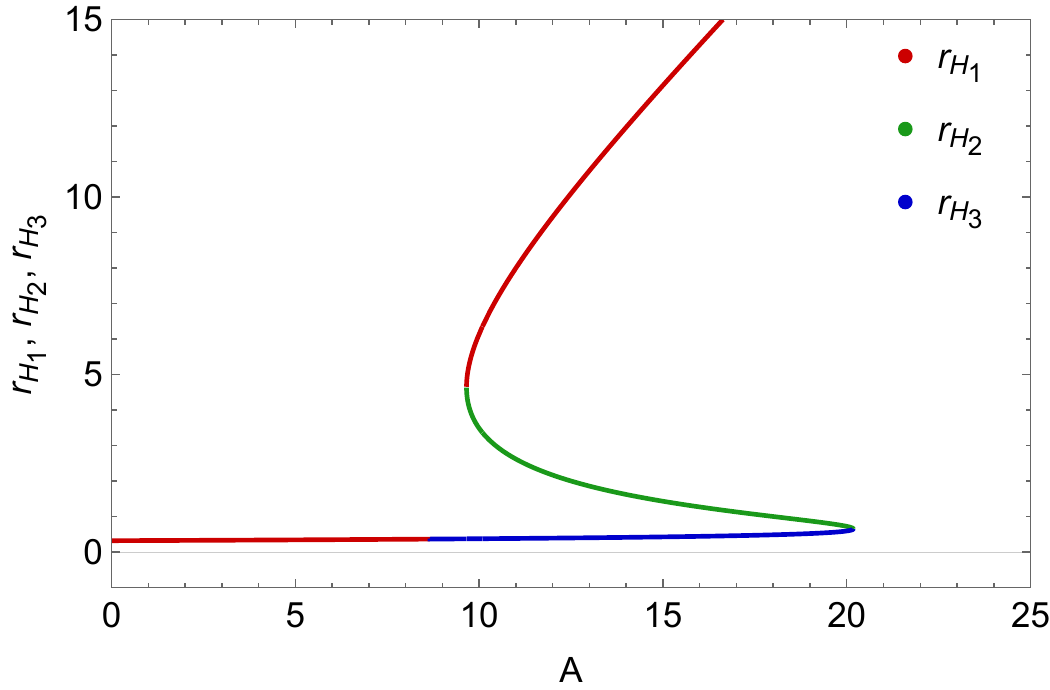}
    \caption{Roots $r_{H_1}$, $r_{H_2}$ and $r_{H_3}$ for $B=25$ and $C=8$. In the range $10<A<20$, where all three roots are positive real and $r_{H_1}>r_{H_2}>r_{H_3}$, the largest root $r_{H_1}$ gives the solution of the event horizon of the black hole. In the rest of the range, where only one root is positive real, it provides the size of the event horizon.}
    \label{Horizons}
\end{figure}

Here, we choose the parametrized static, spherically symmetric, asymptotically flat spacetime to be of the following form
\begin{eqnarray}
ds^2 &=& - f(r) dt^2 + \frac{1}{f(r)} dr^2 + r^2 (d\theta^2+ \sin^2 \theta \, d\phi^2) \label{Metric}
\end{eqnarray}
with metric function
\begin{eqnarray}
    f(r) &=& 1-\frac{A}{r}+\frac{B}{r^2}-\frac{C}{r^3} \label{metric function}
\end{eqnarray}
where the parameters $A,B$ and $C$ are constants. We restrict these constants to be positive definite, which allows three distinct solutions for photon/anti-photon spheres outside the event horizon. The event horizon of the black hole is obtained as the largest solution $r=r_H$ such that $f(r_H)=0$. The above metric function leads to a cubic equation for $f(r)=0$, and the three roots are obtained as
\begin{eqnarray}
r_{H_1} &=& \frac{A}{3} + \frac{\sqrt{Q}}{3} \left[\cos(\xi/3)+\sqrt{3} \sin(\xi/3)\right]\\
r_{H_2} &=& \frac{A}{3} + \frac{\sqrt{Q}}{3}  \left[\cos(\xi/3)-\sqrt{3} \sin(\xi/3)\right]\\
r_{H_3} &=& \frac{A}{3} - \frac{\sqrt{Q}}{3}\left[2\,\cos(\xi/3)\right]
\end{eqnarray}
where,
\begin{eqnarray}
    Q = \left(A^2- 3 B\right) & \text{and} & \xi = \tan^{-1}(y/x)
\end{eqnarray}
with $x = \left(-2A^3+9 A B-27 C\right)$ and $y=\sqrt{4Q^3-x^2}$. In the parameter space of $\{A, B, C\}$ where $y>0$, all three roots are real valued and positive, and they satisfy $r_{H_1}>r_{H_2}>r_{H_3}$. We have plotted the real valued roots against the black hole parameter $A$ in Fig. \ref{Horizons} below with fixed values of $B=25$ and $C=8$. 
In the range $10<A<20$, all three roots are positive real and $r_{H_1}>r_{H_2}>r_{H_3}$. For the parameter space of $\{A, B, C\}$ where $y\leq 0$, two of the roots become complex valued, and only one remains real and positive, providing the location of the event horizon. From Fig. \ref{Horizons}, we can see that the event horizon in this case is given by either $r_{H_1}$ or $r_{H_3}$. In the case where all three roots have positive real values, we consider the largest root given by $r_{H_1}$ as the event horizon.

\subsection{Null Geodesics}\label{Null geodesics}
Since the background spacetime given by the metric, Eq.(\ref{Metric}), is spherically symmetric, it is sufficient to consider the null geodesic motion in the equatorial plane due to angular momentum conservation. Then the equations governing the motion of photons are,
\begin{eqnarray}
    f(r) \,\dot{t} &=& E \\
    r^2 \, \dot{\phi} &=& L\\
    \frac{\dot{r}^2}{L^2}+ V(r) &=& \frac{1}{b^2}
\end{eqnarray}
where $E$ and $L$ are the conserved energy and angular momentum of the photon, and $b=L/E$ is the impact parameter of the photon's trajectory. The effective potential $V(r)$ encountered by the photon along the geodesic is
\begin{eqnarray}
V(r) &=& \frac{f(r)}{r^2}
\end{eqnarray}
For the choice of metric function in Eq.(\ref{metric function}), this potential can have either one or three points of extrema (maxima/minima) depending on the black hole parameters $A, B$, and $C$. These extrema determine the locations of photon/anti-photon spheres in the background spacetime. The radii of photon spheres thus satisfy,
\begin{eqnarray}
    \frac{dV}{dr} &=& \frac{-2 r^3+3 A r^2-4 B r+5 C}{r^6}=0
\end{eqnarray}
The roots of the cubic polynomial in the above equation are given by,
\begin{eqnarray}
r_1 &=& \frac{A}{2} - \frac{\sqrt{M}}{2\times 3^{2/3}} \left[\cos(\eta/3)+\sqrt{3} \sin(\eta/3)\right]\\
r_2 &=& \frac{A}{2} - \frac{\sqrt{M}}{2\times 3^{2/3}} \left[\cos(\eta/3)-\sqrt{3} \sin(\eta/3)\right]\\
r_3 &=& \frac{A}{2} + \frac{\sqrt{M}}{2\times 3^{2/3}} \left[2\,\cos(\eta/3)\right]
\end{eqnarray}
where,
\begin{eqnarray}
M = 3^{1/3} \left(3 A^2- 8 B\right) &\text{and}& \eta = \tan^{-1}(Y/X)
\end{eqnarray}
with $X = 9\left(A^3-4 A B+10 C\right)$ and $Y=\sqrt{M^3-X^2}$. When all three roots are real, they satisfy $r_1<r_2<r_3$. One can verify that, in the case of Schwarzschild spacetime, where $B=C=0$ and $A=2 m$, these three roots reduce to $r_1=r_2=0$ and $r_3=3m$. Further, one can write the following relations for these three roots:
\begin{eqnarray}
    r_1+r_2+r_3 &=& \frac{3A}{2}\label{Vieta's Formula 1}\\
    r_1 r_2+r_2 r_3+r_1 r_3 &=& 2 B\label{Vieta's Formula 2}\\
    r_1 r_2 r_3 &=& \frac{5 C}{2} \label{Vieta's Formula 3}
\end{eqnarray}
We now assess the stability of the roots when all three roots are real. We evaluate the second derivative of the potential,
\begin{eqnarray}
    V''(r)=\frac{d^2 V}{dr^2} &=& \frac{6 r^3-12 A r^2+20 B r-30 C}{r^7}
\end{eqnarray}
at the location of the three roots. The expressions are simplified using Eqs.(\ref{Vieta's Formula 1})-(\ref{Vieta's Formula 3}) to substitute for $A$, $B$, and $C$ in terms of the roots $r_1$, $r_2$, and $r_3$, to get,
\begin{eqnarray}
    \frac{d^2 V}{dr^2}\Bigg|_{r=r_1} &=& \frac{-2}{{r_1}^6} (r_2-r_1)(r_3-r_1) \label{stability of r1}\\
    \frac{d^2 V}{dr^2}\Bigg|_{r=r_2} &=& \frac{2}{{r_2}^6} (r_2-r_1)(r_3-r_2) \label{stability of r2}\\
     \frac{d^2 V}{dr^2}\Bigg|_{r=r_3} &=& \frac{-2}{{r_3}^6} (r_3-r_1)(r_3-r_2) \label{stability of r3}
\end{eqnarray}
Considering the inequalities $r_1<r_2<r_3$, we can conclude that $V''(r)$ at $r_1$ and $r_3$ is negative, thus implying that the unstable photon spheres lie at $r_1$ and $r_3$, whereas $V''(r)$ at $r_2$ is positive and the stable anti-photon sphere lies at $r_2$. This provides the structure of the double-peak potential with two peaks at $r_1$ and $r_3$ and a minimum in between at $r_2$. The photons oscillate about the anti-photon sphere and spiral asymptotically towards the photon spheres. The photons orbiting the photon spheres can escape to a faraway observer, leading to the image of a black hole shadow and photon rings. Depending on the magnitude of the potential at the two photon spheres, the black hole's image will have one or two photon rings. Henceforth, we refer to the two photon spheres at $r_1$ and $r_3$ as inner and outer photon spheres, respectively.

In the spacetimes where the magnitude of the potential at the inner photon sphere is smaller than that at the outer photon sphere, that is, $V(r_1)<V(r_3)$, only one photon ring can be seen in the observed image, and the image will have features similar to those of a Schwarzschild black hole. This has been inferred in the examination of the shadows of different black holes with multiple photon spheres \cite{thin-shell wormholes, Traversable wormholes, hairy Schwarzschild, Hairy RN lensing}. We therefore focus on the case wherein the magnitude of potential at the outer photon sphere is smaller than that at the inner photon sphere, that is $V(r_1)>V(r_3)$. In such a case, the black hole image will be comprised of the central brightness depression along with two separated photon rings. The sizes of these photon rings as seen by an observer at asymptotic infinity are given by the critical impact parameters as,
\begin{eqnarray}
    b_{cr_1}=\frac{1}{\sqrt{V(r_1)}} & \text{and} & b_{cr_3}=\frac{1}{\sqrt{V(r_3)}}
\end{eqnarray}
The images of photon rings and shadows have been examined for many spacetime solutions with multiple photon spheres. These investigations were primarily focused on the observational features of the image and the photon rings and sub-ring structures illuminated by an accretion disk with the observer located on the axis of the disk \cite{thin-shell wormholes, Traversable wormholes, hairy Schwarzschild, Hairy RN lensing}. The influence of an off-axis observer position on the resulting observational features requires further investigation. 

Additionally, the imprints of multiple photon spheres will also manifest in the time-domain analysis of the lensed images. The presence of two unstable photon spheres, combined with $V(r_1)>V(r_3)$, can produce a non-monotonic evolution of higher-order image positions and time delays with respect to the impact parameters. The time delay between successive images then provides a probe of the underlying spacetime geometry.

In the following sections, we choose a set of values for the black hole parameters $\{A,B,C\}$ leading to the double-peak structure of the potential with $V(r_1)>V(r_3)$ and explore the effects of such a potential on the photon's motion and, in turn, on the image positions and time delay.

\section{Double-peak potentials} \label{double peak potentials}

Since the angular sizes of the two photon rings are determined by the magnitudes of the potential at the respective peaks, they are insensitive to the properties of the intervening potential well. In particular, varying the minimum of the potential while keeping the magnitudes at peaks fixed leaves the shadow structure unchanged. As a result, different sets of the parameters $\{A, B, C\}$ can produce identical static shadow images. Time delay observables, however, depend sensitively on the full potential profile and therefore provide a means to break this degeneracy and probe the position and depth of the intermediate potential well.

We now consider two sets of black hole parameters, such that the magnitudes of the potential at the inner photon sphere and the outer photon sphere of set 1 are the same as those of set 2, retaining the inequality $V(r_1)>V(r_3)$. The chosen values of parameters $A$, $B$, and $C$ along with the radii of the event horizon ($r_H$), the photon/anti-photon spheres ($r_1,r_2,r_3$) and the magnitudes of potential at the photon/anti-photon spheres are given below in the Table \ref{BH Parameters}. We have chosen the black hole parameters such that $V(r_1) = 0.020$ and $V(r_3) = 0.014$ for both sets 1 and 2. 
\begin{table}[H]
\centering
\begin{tabular}{|c|c|c|c|c|c|c|c|c|c|c|}
\hline
 & $A$ & $B$ & $C$ & $r_H$ &$r_1$ & $r_2$ & $r_3$ & $V(r_1)$ & $V(r_2)$ & $V(r_3)$\\
  \hline
\textit{BH1} & $4.8$ & $7.81827$ & $4.17736$ & $1.1687$ & $1.34743$ & $2.02487$ & $3.8277$ & $0.020$ & $0.00809$ & $0.014$\\
 \hline
\textit{BH2} & $5$ & $8.84953$ & $5.2493$ & $1.29774$ & $1.54085$ & $2.37877$ & $3.58038$ & $0.020$ & $0.01273$ & $0.014$\\
 \hline
\end{tabular}
\caption{Two sets of black hole parameters with $V(r_1)=0.020$ and $V(r_3)=0.014$ labeled as \textit{BH1} and \textit{BH2}.}
\label{BH Parameters}
\end{table}
We label these black holes with the two sets of parameters as \textit{BH1} and \textit{BH2} for ease of reference. This choice of equal magnitudes of $V(r_1)$ and $V(r_3)$ for \textit{BH1} and \textit{BH2} results in both black holes having the same size of the two photon rings in the black hole images as seen by a faraway observer.

For an observer at $r=r_{obs}$, the angle $\alpha$ between a light ray of impact parameter $b$ and the radial direction pointing towards the black hole is given by \cite{Review}
\begin{eqnarray}
    \sin \alpha &=& \frac{b \,\sqrt{f(r_{obs})}}{r_{obs}} \label{angular distance alpha}
\end{eqnarray}
The observed angular size of the two photon rings, say $\alpha_1$ and $\alpha_3$, will then be,
\begin{eqnarray}
    \alpha_1 &=& \sin^{-1}\left[\frac{b_{cr_1} \,\sqrt{f(r_{obs})}}{r_{obs}}\right]=\sin^{-1}\left[\frac{\sqrt{f(r_{obs})}}{r_{obs}\sqrt{V(r_1)}}\right]\\
    \alpha_3 &=& \sin^{-1}\left[\frac{b_{cr_3} \,\sqrt{f(r_{obs})}}{r_{obs}}\right]=\sin^{-1}\left[\frac{\sqrt{f(r_{obs})}}{r_{obs}\sqrt{V(r_3)}}\right]
\end{eqnarray}
Given that $V(r_1)$ and $V(r_3)$ are equal for both the black holes, the associated angular sizes $\alpha_1$ and $\alpha_2$ of the observed photon rings are the same. Consequently, the photon-ring observations alone will be insufficient to distinguish between the two black holes; as a result, the parameter space $\{A, B, C\}$ remains degenerate. The critical impact parameters corresponding to the two unstable photon spheres of \textit{BH1} and \textit{BH2} are
\begin{eqnarray}
    b_{cr_1} = \frac{1}{\sqrt{V(r_1)}} = 7.07107 \quad & \text{and} & \quad
    b_{cr_3} = \frac{1}{\sqrt{V(r_3)}} = 8.45154
\end{eqnarray}
which determine the sizes of the photon rings as seen by an observer at asymptotic infinity.

\begin{figure}[h]
    \centering
    \includegraphics[scale=0.5]{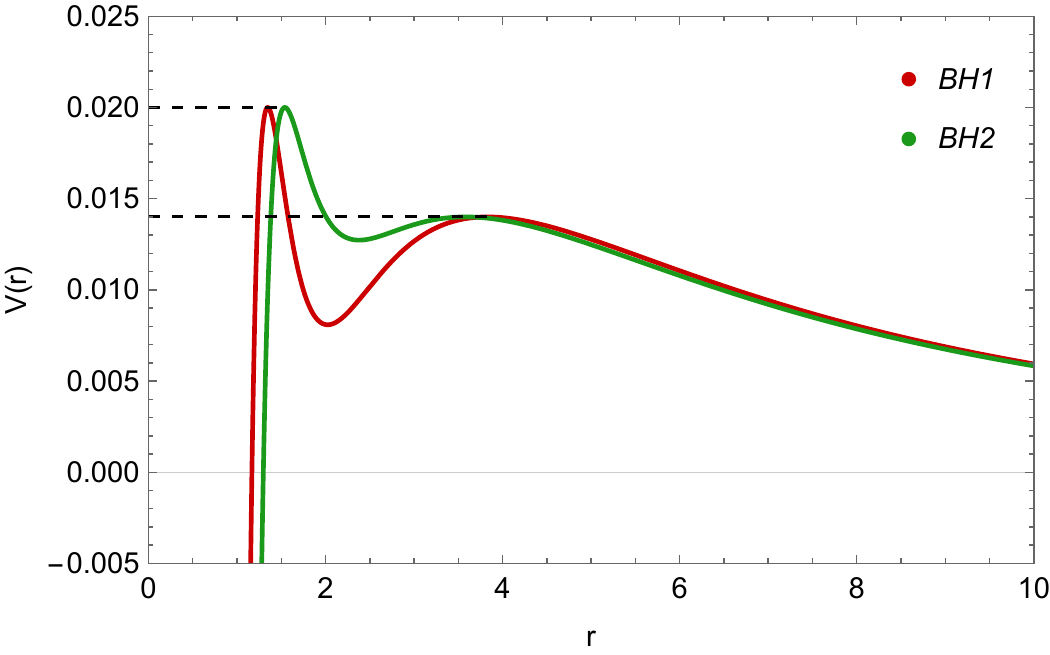}
    \caption{Double-peak potentials for \textit{BH1} (red curve) and  \textit{BH2} (green curve). The black dashed lines denote $V=0.020$ and $V=0.014$.}
    \label{Potentials for 2 sets}
\end{figure}

To illustrate the complete structure of the potential encountered by the photon along its trajectory, we plotted the potential functions $V(r)$ in Fig. \ref{Potentials for 2 sets} for the two black holes \textit{BH1} and \textit{BH2}. From the figure, it is evident that, although the magnitude of potential is equal for both black holes at the two peaks, the locations of the inner and outer photon spheres, as well as the anti-photon sphere, differ in both cases. Furthermore, the magnitude of potential at anti-photon spheres is significantly different for \textit{BH2} compared to \textit{BH1} with $V(r_2)_{BH2}>V(r_2)_{BH1}$. In other words, the potential well near the anti-photon sphere is shallower in \textit{BH2} compared to \textit{BH1}.

In the next sections, we examine the photon trajectories in the background of these two black holes. We use the following equations governing the photon's motion along time ($t$) and angular ($\phi$) coordinates:
\begin{eqnarray}
     \frac{dt}{dr} &=& \pm \frac{1}{b\,f(r)} \, \left( \frac{1}{b^2}-V(r)\right)^{-1/2}\label{t-equation}\\
     \frac{d\phi}{dr} &=& \pm \frac{1}{r^2} \, \left( \frac{1}{b^2}-V(r)\right)^{-1/2}\label{phi-equation}
\end{eqnarray}
where the $+$ and $-$ signs correspond to outgoing and ingoing null geodesics, respectively. For the spacetime metric in Eq.(\ref{Metric}), the integrand in the above expressions has a complicated form. This restricts further analytical investigation, as solutions $t(r)$ and $\phi(r)$ cannot be obtained through symbolic integration. Hence, for the rest of the analysis, we resort to numerical computation.

\section{Comparison with the Schwarzschild spacetime} \label{comparison with SCh}

We first explore the implications of the double-peak potential on null geodesics by comparing it with the Schwarzschild spacetime. We evaluate and compare the angular distance covered by a photon and the time taken by the photon along a trajectory in the background of \textit{BH1}, \textit{BH2}, and the Schwarzschild spacetime. Comparison with the single-peak potential in the Schwarzschild background helps to elucidate the effects of the double-peak potential.

To study the null geodesics, we consider an observer located at $(r_{obs},\theta_{obs},\phi_{obs})=(50,\pi /2,0)$. We then plot the trajectories using the backward ray-tracing method. The trajectories in the \textit{BH1} spacetime are plotted in Fig. \ref{All trajectories} below. The trajectories with $b>b_{cr_3}$ that have turning points outside the outer photon sphere at $r_3$ are shown in Fig. \ref{Set1:Outside PS2}, and trajectories with impact parameter $b_{cr_1}<b<b_{cr_3}$ are shown in Fig. \ref{Set1:Inside PS2}.
\begin{figure}[ht!]
    \centering
     \begin{subfigure}[c]{0.45\textwidth}
          \centering
          \includegraphics[scale=0.90]{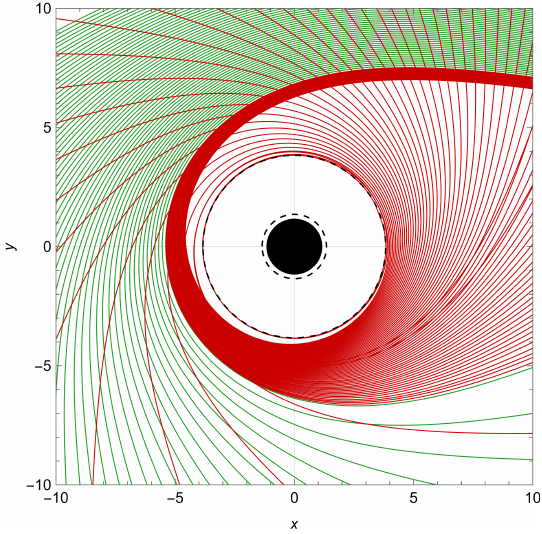}
          \caption{$b>b_{cr_3}$}
          \label{Set1:Outside PS2}
     \end{subfigure}
     \hfill
     \begin{subfigure}[c]{0.45\textwidth}
         \centering
         \includegraphics[scale=0.90]{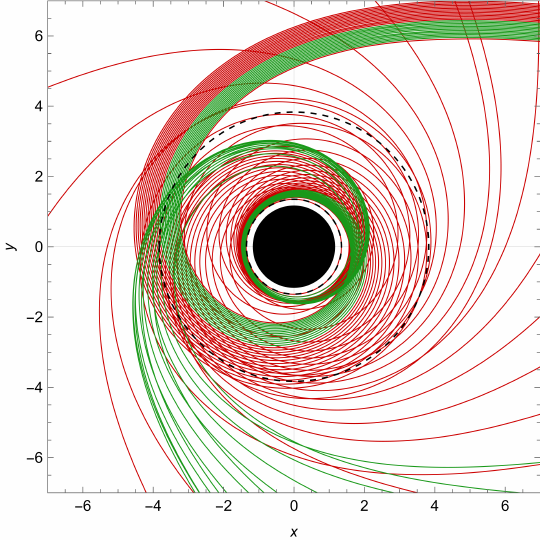}
         \caption{$b_{cr_1}<b<b_{cr_3}$}
         \label{Set1:Inside PS2}
     \end{subfigure}
     \caption{Photon trajectories in the spacetime of \textit{BH1}. The two photon spheres are shown by black dashed circles at $r_1$ and $r_3$, and the central black disc represents the black hole. (a) Trajectories with turning point $r_{turn}>r_3$. The green and red trajectories correspond to $\Delta\phi<2\pi$ and  $\Delta\phi\geq 2\pi$, respectively. (b) Trajectories with turning point $r_1<r_{turn}<r_3$. The green and red trajectories correspond to $\Delta\phi<6\pi$ and  $\Delta\phi\geq 6\pi$, respectively.}
     \label{All trajectories}
\end{figure}

Considering the point sources to be located uniformly on a circle of radius $R_s=15\sqrt{2}$ around the black hole, we evaluated the total angular distance $\Delta\phi$ covered by these trajectories to reach a circle of radius $R_s=15\sqrt{2}$ starting from the observer's location $r_{obs}=50$ as,
\begin{eqnarray}
    \Delta \phi &=& \int_{r_{turn}}^{R_s} \frac{1}{r^2} \, \left( \frac{1}{b^2}-V(r)\right)^{-1/2} dr +\int_{r_{turn}}^{r_{obs}} \frac{1}{r^2} \, \left( \frac{1}{b^2}-V(r)\right)^{-1/2} dr \label{angular distance formula}
\end{eqnarray}
where $r_{turn}$ is the turning point for the trajectory with impact parameter $b$. The green lines in Fig. \ref{Set1:Outside PS2} represent the trajectories with $\Delta\phi<2\pi$, and red lines represent trajectories with $\Delta\phi\geq 2\pi$. In Fig. \ref{Set1:Inside PS2}, the green lines represent trajectories with $\Delta\phi<6\pi$ and the red lines represent trajectories with $\Delta\phi\geq 6\pi$. One can note here that, unlike Fig. \ref{Set1:Outside PS2}, the red trajectories in Fig. \ref{Set1:Inside PS2} appear with impact parameters both less than and greater than those of green trajectories.

We also evaluated $\Delta\phi$ for photons in \textit{BH2} spacetime and the Schwarzschild spacetime where the mass of the black hole $m$ is chosen such that the critical impact parameter $b_{cr}=3\sqrt{3}\, m=b_{cr_1}$. We have plotted the angular distance covered, $\Delta\phi$, for impact parameters in the range $(b_{cr_1},15)$ in Fig. \ref{DphiAll} below.

\begin{figure}[ht]
    \centering
    \includegraphics[width=0.8\linewidth]{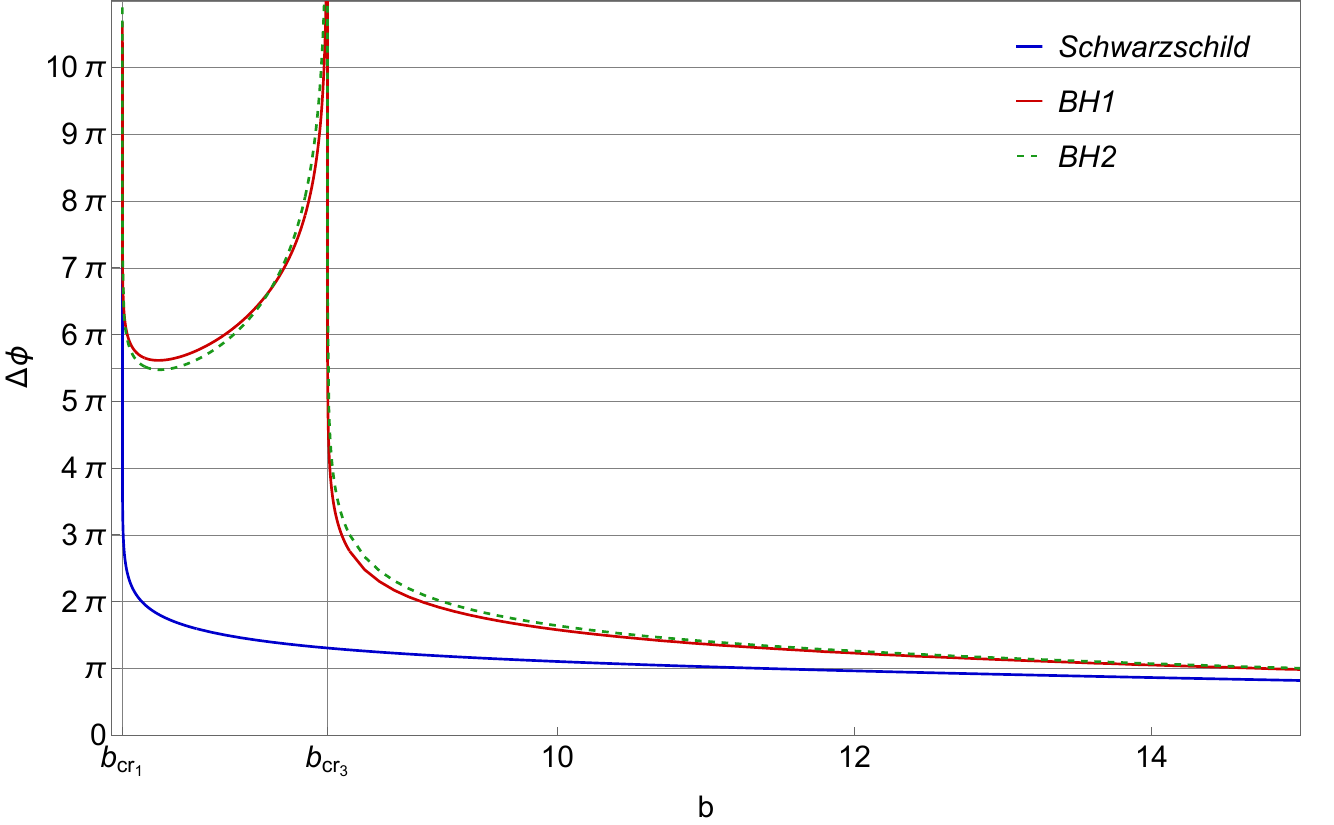}
    \caption{The angular distance $\Delta\phi$ covered by the photon trajectories starting from the observer's location at $r_{obs}=50$ to reach a circle of sources at radius $R_s=15\sqrt{2}$. In contrast to Schwarzschild spacetime, where $\Delta\phi$ approaches to infinity only close to $b_{cr_1}$, such behavior occurs additionally near $b_{cr_3}$ from both directions $b<b_{cr_3}$ and $b>b_{cr_3}$ for \textit{BH1} and \textit{BH2}.}
    \label{DphiAll}
\end{figure}

From the figure, we infer the following conclusions: (i) For \textit{BH1} and \textit{BH2}, $\Delta\phi$ decreases monotonically for $b>b_{cr_3}$, that is, for trajectories outside the outer photon sphere at $r_3$, similar to the case of the Schwarzschild black hole. However, between the two photon spheres at $r_1$ and $r_3$, $\Delta\phi$ decreases to a minimum value between $5\pi$ and $6\pi$ and further increases, approaching large values close to the outer photon sphere. This implies that the photon trajectories outside the outer photon sphere with $b>b_{cr_3}$ can complete any $n$ number of half ($\pi$) turns around the black hole, starting from $n=0$, as in the case of the Schwarzschild black hole. However, the trajectories falling inside the outer photon sphere with $b_{cr_1}<b<b_{cr_3}$, complete at least a nonzero minimum number of turns, say $N_{min}$, around the black hole before reaching the observer. From Fig. \ref{DphiAll}, we note that for both \textit{BH1} and \textit{BH2}, $N_{min}=5$. (ii) The minimum value of $\Delta\phi$, say $\phi_{min}$, is smaller for \textit{BH2} than for \textit{BH1}. More precisely, $\phi_{min}<5\pi+\pi/2$ for \textit{BH2} and $\phi_{min}>5\pi+\pi/2$ for \textit{BH1}. If the impact parameter corresponding to $\phi_{min}$ is $b_{min}$, then the trajectories with the impact parameter close to $b_{min}$ will cover a smaller angular distance $\Delta\phi$ in \textit{BH2} spacetime compared to that in \textit{BH1} spacetime. This is expected to be the combined effect of a larger turning point $r_{turn}$ and shallower potential well in \textit{BH2} compared to \textit{BH1}. (iii) Unlike the Schwarzschild black hole, there exist three different photon trajectories that cover equal angular distance $\Delta\phi>\phi_{min}$ corresponding to three different impact parameters in the \textit{BH1} and \textit{BH2} spacetimes. (iv) A large fraction of the range of impact parameter between $b_{cr_1}$ and $b_{cr_3}$ is covered by trajectories with $5\pi<\Delta\phi \leq 8 \pi$.

We further evaluated the time taken by these trajectories to reach the observer at $r_{obs}$ as,
\begin{eqnarray}
    T_{obs} &=& \int_{r_{turn}}^{R_s} \frac{1}{b\,f(r)} \, \left( \frac{1}{b^2}-V(r)\right)^{-1/2} dr +\int_{r_{turn}}^{r_{obs}}\frac{1}{b\,f(r)} \, \left( \frac{1}{b^2}-V(r)\right)^{-1/2} dr \label{Time of observation}
\end{eqnarray}
We plotted this time below in Fig. \ref{TobsAll} for the same range of impact parameters as in Fig. \ref{DphiAll}.
\begin{figure}[ht]
    \centering
    \includegraphics[width=0.8\linewidth]{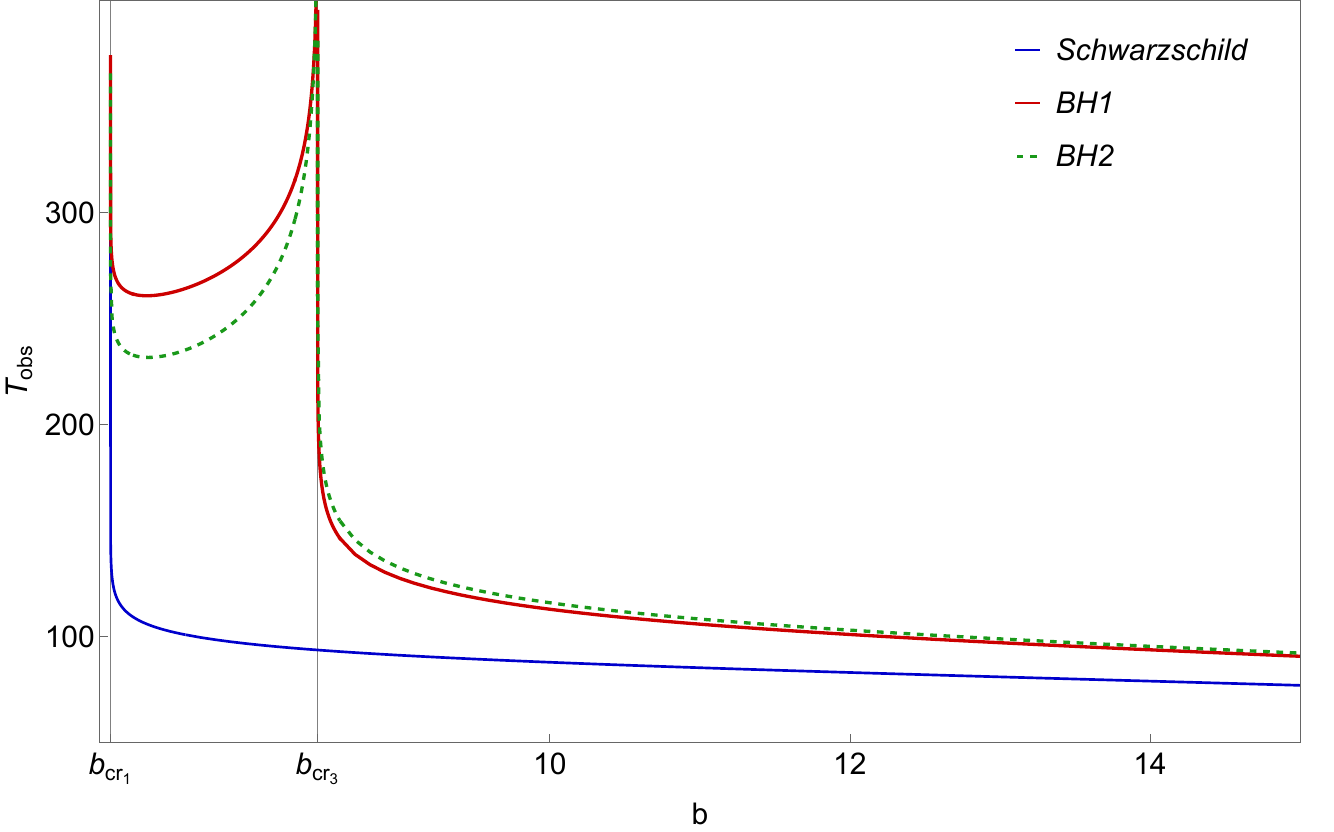}
    \caption{The time taken $T_{obs}$ covered by the photon trajectories starting from the observer's location at $r_{obs}=50$ to reach a circle of sources at radius $R_s=15\sqrt{2}$. As compared to the angular distance $\Delta\phi$ in Fig. \ref{DphiAll}, the difference in $T_{obs}$ for \textit{BH1} and \textit{BH2} is more prominent.}
    \label{TobsAll}
\end{figure}
For the two black holes \textit{BH1} and \textit{BH2}, though the difference between the angular distance covered $\Delta\phi$ is small close to the impact parameter $b_{min}$, the corresponding time $T_{obs}$ differs by a considerable amount. From the figure, it is evident that if $T_{min}$ is the minimum value of the $T_{obs}$ curve between $b_{cr_1}$ and $b_{cr_3}$, then for $T_{obs}>T_{min}$, three distinct images of the source can be observed simultaneously. Moreover, for any chosen impact parameter $b$ within the interval $b_{cr_1}<b<b_{cr_3}$, the photon in \textit{BH2} spacetime requires less time to reach the observer than in \textit{BH1} spacetime. This indicates that, for the black hole spacetime constructed in section \ref{Spacetime model}, the images in the region between the two photon rings will appear earlier if the potential well is shallower.

The minimum values $T_{min}$ and $\phi_{min}$ can be shown to occur at the same impact parameter $b$ as demonstrated below. It is important here to note that the expressions of $\Delta\phi$ and $T_{obs}$ are improper integrals as the integrand diverges near the lower limit $r_{turn}$ of integration. Such integrals are evaluated by replacing the integration limit by $r_{turn}+\epsilon$ and then taking the limit $\epsilon\to 0$. The resulting expressions for $\Delta\phi$ and $T_{obs}$ are given by
\begin{eqnarray}
\Delta \phi &=& \lim_{\epsilon \to 0} \left(\,\int_{r_{turn}+\epsilon}^{R_s}+\int_{r_{turn}+\epsilon}^{r_{obs}}\, \right) \left[
\frac{1}{r^2} \, \left( \frac{1}{b^2}-V(r)\right)^{-1/2} dr  \right]\\
T_{obs} &=& \lim_{\epsilon \to 0} \left(\,\int_{r_{turn}+\epsilon}^{R_s}+\int_{r_{turn}+\epsilon}^{r_{obs}}\, \right) \left[ \frac{1}{b\,f(r)} \, \left( \frac{1}{b^2}-V(r)\right)^{-1/2} dr \right]
\end{eqnarray}
We now differentiate these quantities with respect to the impact parameter $b$. Since the turning point $r_{turn}$ depends on the impact parameter, the derivatives are evaluated using the Leibniz integral rule as,
\begin{eqnarray}
\frac{d\Delta\phi}{db} &=& \lim_{\epsilon \to 0}\, \Bigg[\, \left(\,\int_{r_{turn}+\epsilon}^{R_s}+\int_{r_{turn}+\epsilon}^{r_{obs}}\, \right) \frac{1}{r^2}  \frac{d}{db} \left\{\left( \frac{1}{b^2} - V(r) \right)^{-1/2} \right\} dr\, \nonumber\\
& & \qquad - 2\,\left[ \,\frac{1}{r^2} \left( \frac{1}{b^2} - V(r) \right)^{-1/2} \right]_{r_{turn}+\epsilon}\, \frac{d\,r_{turn}}{db} \,\Bigg] \label{dphi/db equation}\\
\frac{dT_{obs}}{db} &=& \lim_{\epsilon \to 0}\, \Bigg[\, \left(\,\int_{r_{turn}+\epsilon}^{R_s}+\int_{r_{turn}+\epsilon}^{r_{obs}}\, \right) \frac{1}{f(r)}\frac{d}{db}  \left\{\frac{1}{b} \left( \frac{1}{b^2} - V(r) \right)^{-1/2} \right\} dr \nonumber\\
& &\qquad - 2\,\left[ \frac{1}{b f(r)} \left( \frac{1}{b^2} - V(r) \right)^{-1/2} \right]_{r_{turn}+\epsilon} \frac{d\,r_{turn}}{db} \,\Bigg] \label{dT/db equation}
\end{eqnarray}
Using the relation
\begin{eqnarray}
\frac{d}{db}  \left\{\frac{1}{b} \left( \frac{1}{b^2} - V(r) \right)^{-1/2} \right\} &=& b\, V(r)\, \frac{d}{db} \left( \frac{1}{b^2} - V(r) \right)^{-1/2}
\end{eqnarray}
along with $V(r)=f(r)/r^2$ to rewrite the integrand in Eq.(\ref{dT/db equation}), we get 
\begin{eqnarray}
\frac{dT_{obs}}{db} &=& b\; \lim_{\epsilon \to 0}\, \Bigg[\, \left(\,\int_{r_{turn}+\epsilon}^{R_s}+\int_{r_{turn}+\epsilon}^{r_{obs}}\, \right) \frac{1}{r^2}\frac{d}{db}  \left\{ \left( \frac{1}{b^2} - V(r) \right)^{-1/2} \right\} dr \nonumber\\
& & \qquad \quad - \,\frac{2}{b^2\,V(r_{turn}+\epsilon)} \left[ \frac{1}{r^2 } \left( \frac{1}{b^2} - V(r) \right)^{-1/2} \right]_{r_{turn}+\epsilon} \frac{d\,r_{turn}}{db} \Biggr] \label{dT/db equation 2}
\end{eqnarray}
At the turning point, $b^2 \,V(r_{turn})=1$, thus reducing the last term in the above equation to that in Eq.(\ref{dphi/db equation}) in the limit $\epsilon\to 0$. We can then write
\begin{eqnarray}
    \frac{d \Delta\phi}{d b} &=& \frac{1}{b} \frac{d T_{obs}}{d b}
\end{eqnarray}
Since the extrema occur at values of $b$ for which the derivatives vanish, it follows that the same impact parameter $b_{min}$ is associated with both $T_{min}$ and $\phi_{min}$.

We point out here that though the last term in Eqs. (\ref{dphi/db equation}) and (\ref{dT/db equation 2}) diverges in the limit $\epsilon\to 0$, the entire expressions remain finite everywhere except near the two photon spheres, which is evident from the Figs. \ref{DphiAll} and \ref{TobsAll}. The finiteness of the derivatives can be checked by substituting
\begin{eqnarray}
    \frac{d}{db} \left( \frac{1}{b^2} - V(r) \right)^{-\frac{1}{2}} &=& \frac{2}{b^3 V'(r)} \,\frac{d}{dr} \left( \frac{1}{b^2} - V(r) \right)^{-\frac{1}{2}}
\end{eqnarray}
and 
\begin{eqnarray}
    \frac{dr_{turn}}{db} &=& \frac{-2}{b^3 V'(r_{turn})}
\end{eqnarray}
 in Eqs. (\ref{dphi/db equation}) and (\ref{dT/db equation 2}) and then performing the integration by parts to get
 \begin{eqnarray}
    \int \frac{1}{r^2 V'(r)}\left\{ \frac{d}{dr} \left( \frac{1}{b^2} - V(r) \right)^{-\frac{1}{2}} \right\} dr &=& \frac{1}{r^2 V'(r)} \left( \frac{1}{b^2} - V(r) \right)^{-\frac{1}{2}} \nonumber\\
    & & \;- \int \frac{d}{dr}\left( \frac{1}{r^2 V'(r)} \right) \left( \frac{1}{b^2} - V(r) \right)^{-\frac{1}{2}} dr \qquad
 \end{eqnarray}
The first term on the right-hand side, evaluated at the lower limit ($r_{turn}+\epsilon$) of the two integrations, exactly cancels the divergent term in the limit $\epsilon\to 0$, and the derivatives remain finite.

In the next section, we consider a fixed point source and examine in detail the photon trajectories and time delays caused by the double-peak potential in the background of \textit{BH1} and \textit{BH2} spacetimes.

\section{Time delay of light echoes}\label{Time delay of BHs}

At the location of the observer, the only physical observables are the distance $\alpha$ of the source image from the line of sight, which is a measure of the impact parameter of the photon’s trajectory (as shown in Eq.(\ref{angular distance alpha})), and the time of observation $T_{obs}$. To study the black hole shadow, the light source is usually assumed to be continuous, such as an accretion disk. However, a transient event in the accretion disk could cause a flash of light, thus leading to the light echoes, i.e., multiple images of the flash observed at different impact parameters at different times. These time-delayed images provide additional information about the background geometry, which is necessary to break the degeneracy in the black hole parameters, such as in \textit{BH1} and \textit{BH2}, caused by the equal sizes of photon rings.

We consider a point source and an observer located in the equatorial plane ($\theta=\pi/2$) and study the null geodesics connecting the source to the observer in the spacetimes of \textit{BH1} and \textit{BH2}. The source is chosen to be located at $(R_s,\phi_s)=(15\sqrt{2},\,3\pi/4)$, while the observer is positioned at $(r_{obs},\phi_{obs})=(50,\,0)$. With these locations held fixed, the null geodesics connecting them are determined using a backward ray-tracing method. Each trajectory is labeled by its corresponding impact parameter. The resulting trajectories are plotted in the coordinate plane $(x,y)$ defined by $x= r\cos\phi$ and $y=r\sin\phi$, where the observer is located at $(50,0)$ and the source at $(-15,15)$.

In this setup, we have calculated the following quantities for each trajectory in the background of the two black holes:
\begin{enumerate}
\item The turning point $r_{turn}$ of the trajectory is evaluated such that,
\begin{eqnarray}
    \frac{1}{b^2}-V(r_{turn}) &=& 0
\end{eqnarray}

\item The total angular distance $\Delta\phi$ covered by the photon along the path from source to observer is calculated using Eq.(\ref{angular distance formula}).

\item Considering a flash of light emitted by the source at time $t_s=0$, the time taken by the photon to reach the observer is evaluated using Eq.(\ref{Time of observation}) and it is noted as the time of observation $T_{obs}$.

\item The total number $n$ of half-orbits completed by the photon around the black hole is calculated as $In[\Delta\phi/\pi]$, where $In[x]$ denotes the integer part of $x$. The number $n$ is also referred to as the order of the image.

\item For the photon trajectories falling inside the outer photon sphere with impact parameter $b_{cr_1}<b<b_{cr_3}$, the total number of turns is divided into two parts, namely the turns completed outside the outer photon sphere $n_{out}$, and the turns completed between the two photon spheres $n_{in}$. Since the backward ray-tracing method is used, starting from the observer's location, $n_{out}$ is computed as,
\begin{eqnarray}
    n_{out} &=& In\left[\,\phi_{in}(r_3)-\phi_{obs}\,\right]\, + \, In \left[\,n\pi-\phi_{out}(r_3)\,\right]
\end{eqnarray}
where $\phi_{in}(r)$ and $\phi_{out}(r)$ correspond to the angular distance $\phi(r)$ calculated for the ingoing trajectory from the observer to the turning point and the outgoing trajectory from the turning point to the source, respectively. Once $n_{out}$ is determined, the corresponding value of $n_{in}$ is calculated as $n_{in}=n-n_{out}$.
\end{enumerate}

All these quantities, together with the corresponding impact parameters of the trajectories, are listed in Tables \ref{Set1 Data} and \ref{Set2 Data} for \textit{BH1} and \textit{BH2}, respectively. As evident from Fig. \ref{DphiAll}, multiple trajectories are obtained for $n\geq 6$ with different distributions among $n_{out}$ and $n_{in}$.

\subsection{\textit{BH 1}}

\begin{figure}[h!]
    \centering
    \includegraphics[width=\linewidth]{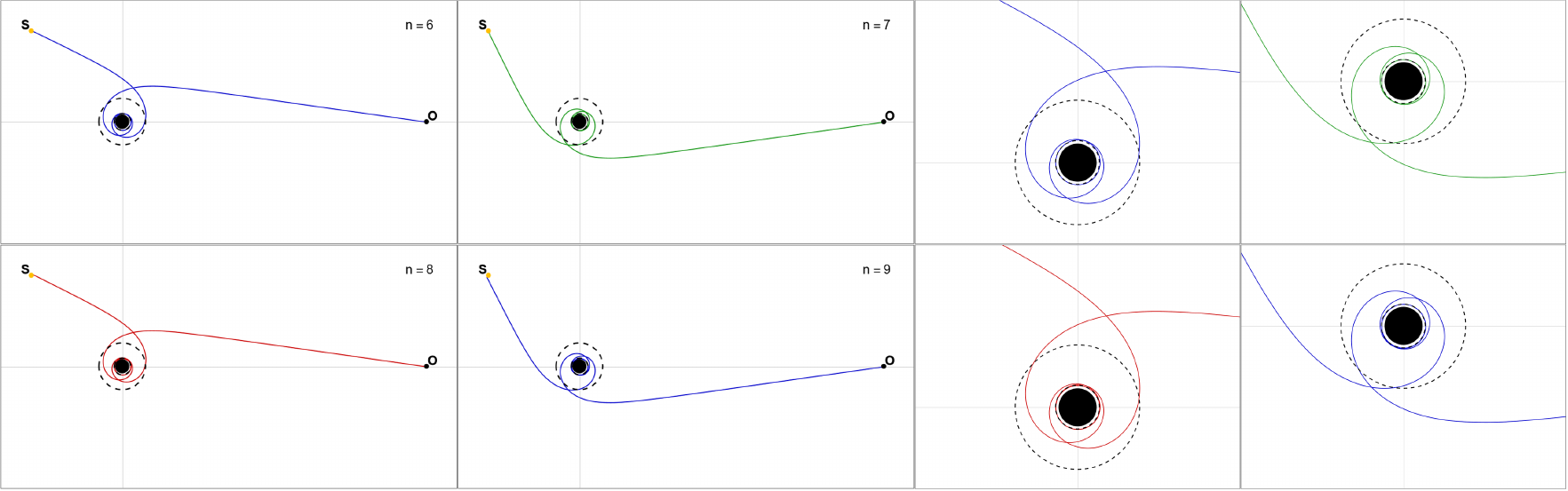}
    \caption{Trajectories with impact parameter close to $b_{cr_1}$ and $n_{out}=0$. The first two columns show the entire trajectory from source(\textbf{S}) to observer(\textbf{O}), while the last two show the orbits close to the inner photon sphere in the same order.}
    \label{Set1:Close to r1}
\end{figure}

\begin{figure}[h!]
    \centering
    \includegraphics  [width=\linewidth]{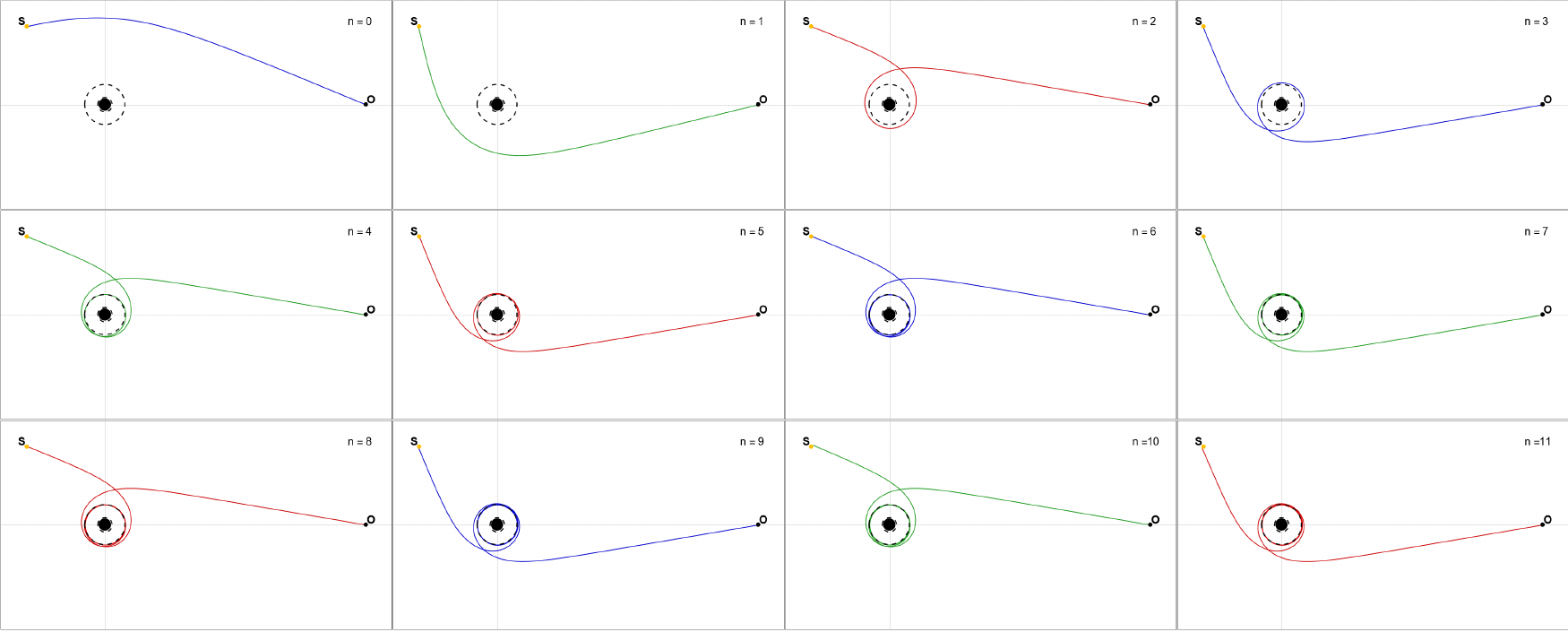}
    \caption{Trajectories with impact parameter $b>b_{cr_3}$ and $n_{in}=0$. The source and observer are represented by \textbf{S} and \textbf{O} respectively.}
    \label{Set1:Outside r3}
\end{figure}

\begin{figure}[h!]
    \centering
    \includegraphics[width=\linewidth]{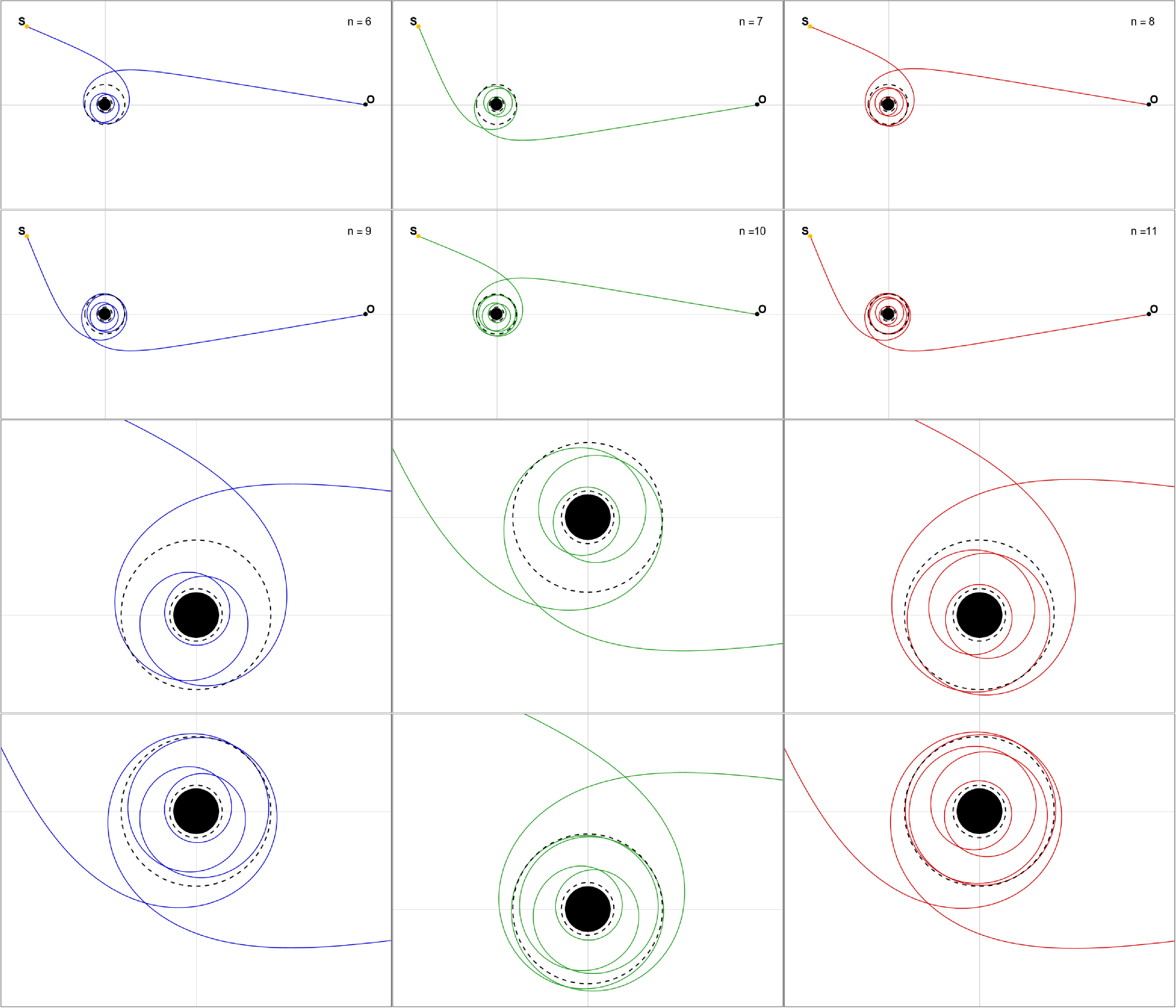}
    \caption{Trajectories with impact parameter $b_{cr_1}<b<b_{cr_3}$ with total number of half-orbits $n$ distributed between $n_{out}$ and $n_{in}$. The first two rows of the figure show the entire trajectory from source \textbf{S} to observer \textbf{O}. The last two rows show the orbits of each of those trajectories close to the photon spheres, in the same order.}
    \label{Set1:Between r1 and r3}
\end{figure}

In the background of \textit{BH1}, we identified the strongly lensed trajectories connecting the source and observer up to $n=11$. We further calculated the quantities mentioned above for these trajectories and have tabulated them in Table \ref{Set1 Data}. To illustrate the exact path followed by the photon, we have plotted these trajectories in Figs. \ref{Set1:Close to r1}, \ref{Set1:Outside r3} and \ref{Set1:Between r1 and r3}.

\begin{table}[h!]
    \centering
    \begin{tabular}{|c|c|c|c|c|c|c|c|c|}
    \hline
    \multicolumn{8}{|c|}{\textit{BH1}} \\
    \hline
    No. & $b$ & $r_{turn}$ & $\Delta \phi$ & $T_{obs}$ & $n$ & $ n_{out} $& $n_{in}$ \\
    \hline
    1 & 7.0710694 & 1.34761 & 28.9932 & 340.214 & 9 & 0 & 9 \\
    \hline
    2 & 7.0710751 & 1.34782 & 27.4434 & 329.589 & 8 & 0 & 8 \\
    \hline
    3 & 7.0717465 & 1.35116 & 22.7751 & 297.312 & 7 & 0 & 7 \\
    \hline
    4 & 7.074174 & 1.35548 & 21.2047 & 286.183 & 6 & 0& 6 \\
    \hline
    5 & 8.09083 & 1.53653 & 21.2054 & 288.762 & 6 & 1 & 5 \\
    \hline
    6 & 8.21458 & 1.55086 & 22.7762 & 301.575 &  7 & 1 & 6 \\
    \hline
    7 & 8.38476 & 1.57003 & 27.4891 & 340.771 & 8 & 1 & 7 \\
    \hline
    8 & 8.40795 & 1.5726 & 29.0597 & 353.96 & 9 & 2 & 7 \\
    \hline
    9 & 8.439535 & 1.5761 & 33.7718 & 393.669 & 10 & 3 & 7 \\
    \hline
    10 & 8.443746 & 1.57656 & 35.3425 & 406.929 & 11 & 3 & 8 \\
    \hline
    11 & 8.4515426 & 3.82806 & 35.3429 & 371.959 & 11 & 11 & 0 \\
    \hline
    12 & 8.4515427 & 3.82825 & 33.7673 & 358.671 & 10 & 10 & 0 \\
    \hline
    13 & 8.4515443 & 3.82974 & 29.0581 & 318.932 & 9 & 9 & 0 \\
    \hline
    14 & 8.4515468 & 3.83085 & 27.4921 & 305.71 & 8 & 8 & 0 \\
    \hline
    15 & 8.4516003 & 3.83932 & 22.7758 & 265.876 & 7 & 7 & 0 \\
    \hline
    16 & 8.4516801 & 3.84566 & 21.2046 & 252.608 & 6 & 6 & 0 \\
    \hline
    17 & 8.4534 & 3.89437 & 16.4955 & 212.811 & 5 & 5 & 0 \\
    \hline
    18 & 8.45598 & 3.93151 & 14.9225 & 199.514 & 4 & 4 & 0 \\
    \hline
    19 & 8.51321 & 4.23739 & 10.2043 & 159.584 & 3 & 3 & 0 \\
    \hline
    20 & 8.60343 & 4.5 & 8.63941 & 146.159 & 2 & 2 & 0 \\
    \hline
    21 & 11.822 & 8.74395 & 3.91976 & 101.794 & 1 &  1 & 0 \\
    \hline
    22 & 19.3459 & 16.6272 & 2.35619 & 78.2723 & 0 & 0 & 0 \\
    \hline
    \end{tabular}
    \caption{Impact parameters $b$ of photon trajectories connecting the source and observer in \textit{BH1} spacetime along with the turning point $r_{turn}$, angular distance covered $\Delta \phi$, time of observation $T_{obs}$, total number of half-orbits $n$, and the corresponding values of $n_{out}$ and $n_{in}$.}
    \label{Set1 Data}
\end{table}

\begin{figure}[h!]
    \centering
    \includegraphics[width=\linewidth]{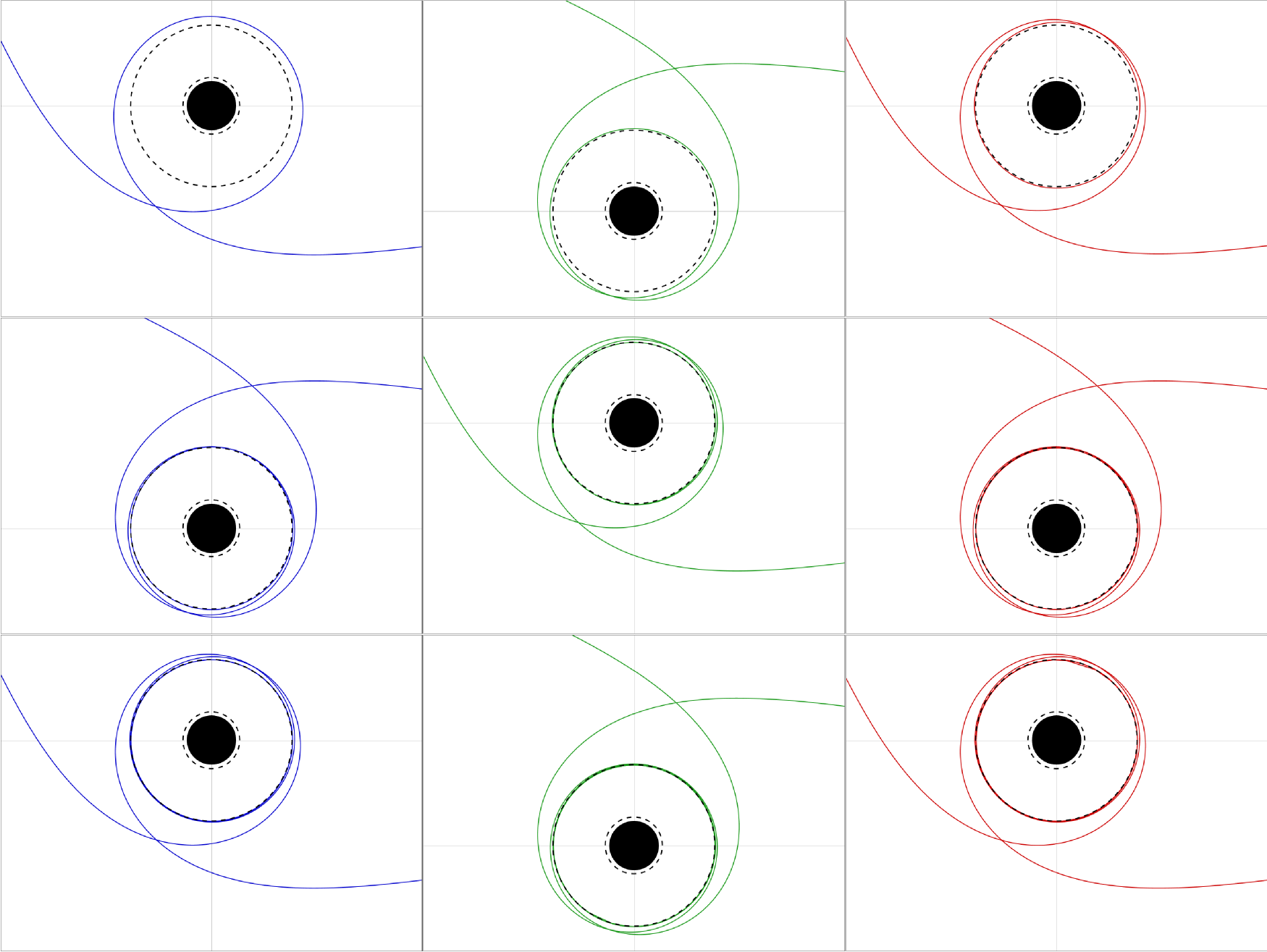}
    \caption{Orbits of the trajectories with $b>b_{cr_3}$ near the outer photon sphere starting from $n=3$ to $n=11$ showing finer details.}
    \label{Set1:Outside r3 magnified}
\end{figure}

Using the data in Table \ref{Set1 Data}, we infer the following key findings:

\begin{enumerate}

\item The trajectories numbered $1-4$ have impact parameter $b$ close to $b_{cr_1}$ and turning point $r_{turn}$ close to $r_1=1.34743$. These trajectories fall directly towards the inner photon sphere and orbit only the inner photon sphere before escaping back to the observer. These trajectories generate the sub-ring structure of the inner photon ring in the black hole image. However, unlike the case of a single-peak potential, the sequence of sub-ring images starts from order $n=6$. The detailed structure of the trajectories is shown in Fig. \ref{Set1:Close to r1}.

\item The trajectories numbered $11-22$ have impact parameters $b>b_{cr_3}$ and orbit only outside the outer photon sphere at $r_3=3.8277$. These trajectories contribute to the sub-ring structure of the outer photon ring in the black hole image. The complete trajectories are plotted in Fig. \ref{Set1:Outside r3} while a magnified view of orbits near the outer photon sphere is presented in Fig. \ref{Set1:Outside r3 magnified} to highlight finer details.

\item The trajectories numbered $5-10$ have impact parameters between the two critical values $b_{cr_1}<b<b_{cr_3}$. These trajectories encircle both the inner and outer photon spheres. However, near the outer photon sphere, photon orbits occur in both the interior and exterior regions (relative to the outer photon sphere), whereas the orbits are confined exclusively to the exterior region in the vicinity of the inner photon sphere, as illustrated in Fig. \ref{Set1:Between r1 and r3}.

\item There are three distinct trajectories with the same order $n$ for $n\geq 6$. We refer to these trajectories as triplets of order $n$. The division of their orbits outside the outer photon sphere $n_{out}$ and between the two photon spheres $n_{in}$ explains the nature of these trajectories. For example, photons with $n=9$, completing $9$ half-orbits around the black hole, follow three different trajectories. The two possibilities are trajectories with all $9$ orbits infinitesimally close to either the inner photon sphere or the outer photon sphere, similar to the unstable photon sphere of the Schwarzschild black hole. Additionally, for the chosen locations of source and observer, a photon can complete $7$ orbits between the two photon spheres and $2$ outside the outer photon sphere. This distribution among $n_{out}$ and $n_{in}$ is expected to change for different relative positions of source and observer.

\item As expected from Fig. \ref{DphiAll}, there are no trajectories with impact parameter $b_{cr_1}<b<b_{cr_3}$ and $n\leq 5$. The minimum number of turns, $n=6$, can be explained from the corresponding trajectory in Fig. \ref{Set1:Between r1 and r3}. For the ingoing part of the trajectory, the first half-turn occurs outside the outer photon sphere, the second is completed during the fall from the outer to the inner photon sphere, and the third close to the inner photon sphere. Owing to the symmetry of the trajectory about the turning point, the remaining three half-turns are covered along the outgoing trajectory. This is the simplest path a trajectory can take with impact parameter $b_{cr_1}<b<b_{cr_3}$ while having nonzero $n_{out}$ and $n_{in}$. As a result, the trajectories with impact parameter $b_{cr_1}<b<b_{cr_3}$ will cover at least $6$ half-turns, if the angular separation between source and observer is greater than $\pi/2$.

\item To visualize the expected form of the observed data, we have plotted the time of observation $T_{obs}$ against the impact parameter $b$ in Fig. \ref{Set1: Tand phi vs b} below. We also plotted the angular distance covered $\Delta\phi$ along with the data of Figs. \ref{DphiAll} and \ref{TobsAll} for completeness.
\begin{figure}[h!]
    \centering
    \begin{subfigure}[c]{0.45\textwidth}
         \centering
         \includegraphics[scale=0.55]{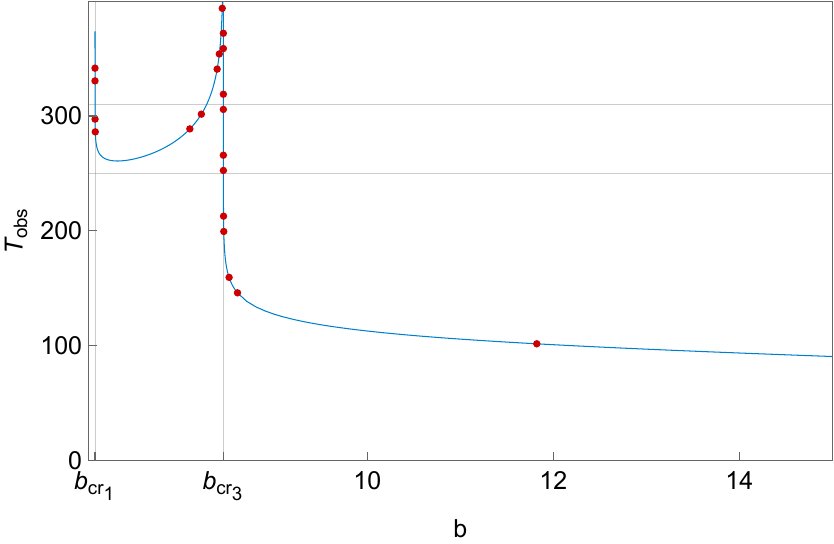}
         \caption{}
         \label{Set1: Tvsb}
     \end{subfigure}
     \hfill
    \begin{subfigure}[c]{0.45\textwidth}
          \centering
          \includegraphics[scale=0.55]{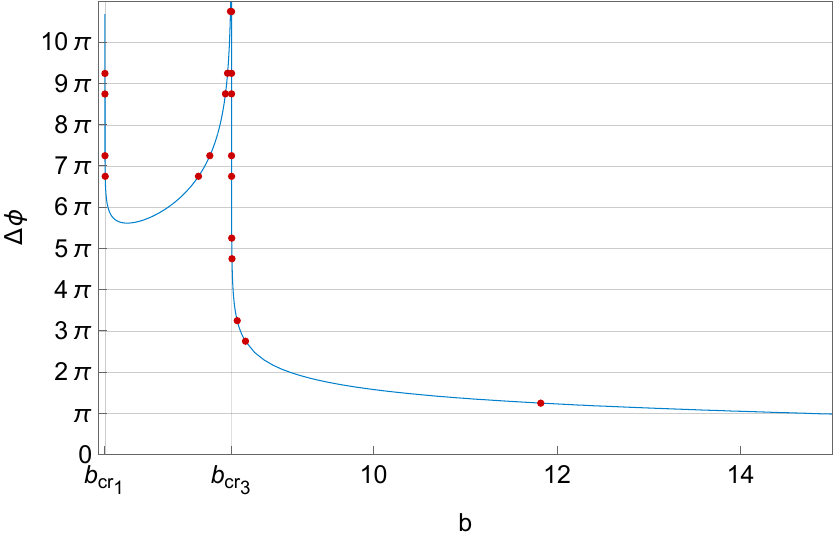}
          \caption{}
          \label{Set1: Phivsb}
     \end{subfigure}
     \caption{The observation time $T_{obs}$ and the angular distance covered $\Delta\phi$ plotted for all trajectories connecting the source and observer. The data from Table \ref{Set1 Data} is shown as red points. The blue line shows the data plotted in Figs. \ref{DphiAll} and \ref{TobsAll} for \textit{BH1} where the sources were uniformly distributed on a circle of radius $R_s$. The two horizontal lines in the left figure denote $T_{obs}=250$ and $T_{obs}=310$.}
     \label{Set1: Tand phi vs b}
\end{figure}
The red points represent the discrete trajectories connecting the source and observer, and they follow the continuous trend shown by the solid curves. The pairs of red points correspond to images appearing on opposite sides of the black hole. Notably, the time difference within each pair is smaller than the delay of the higher-order images. 

\item The images corresponding to $n=6$ and $n=7$ half-turns appear within the time interval $250<T_{obs}<310$. Thus, within a fixed time window, one can observe one $6^{th}$ order image between the two photon rings, in addition to the two $6^{th}$ order images contributing to the sub-ring structure of each photon ring. This implies that an appropriately chosen observation time window can capture the triplets of the same order at different impact parameters.

\item From the plots in Fig. \ref{Set1: Tand phi vs b}, one can note that three trajectories covering the same angular distance $\Delta\phi$ reach the observer at different times. To elucidate the relationship between the angular distance covered and the corresponding observation time for these images, we plot the observation time $T_{obs}$ as a function of $\Delta\phi$ in Fig. \ref{Set1: TvsPhi}. 
\begin{figure}[h!]
    \centering
    \includegraphics[scale=0.60]{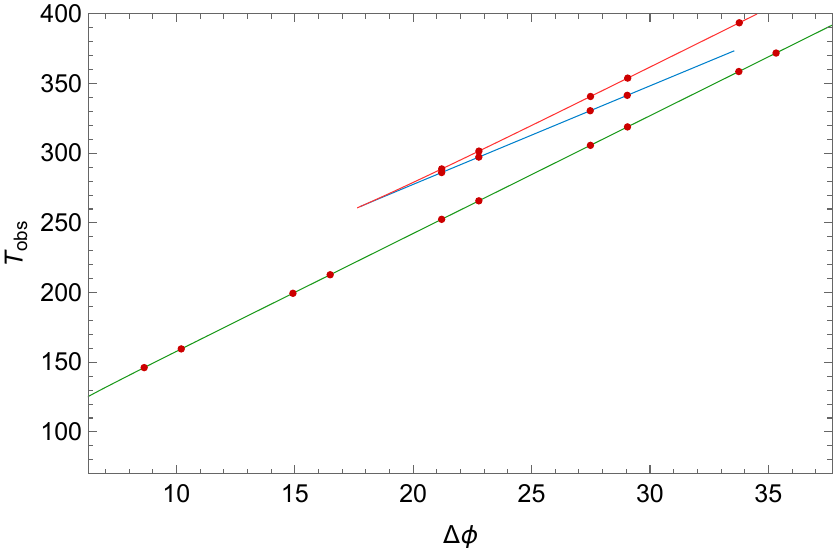}
    \caption{The red points represent the data in Table \ref{Set2 Data}. The three lines correspond to the data for \textit{BH1} in Figs. \ref{DphiAll} and \ref{TobsAll} classified as: (i) Green line: Trajectories outside the outer photon sphere with $b>b_{cr_3}$, (ii) Blue line: Trajectories close to the inner photon sphere with $b_{cr_1}<b\leq b_{min}$, (iii) Red line: Trajectories falling between the two photon spheres with $b_{min}<b<b_{cr_3}$.}
    \label{Set1: TvsPhi}
\end{figure}
The red points represent the data in Table \ref{Set1 Data}. The three lines correspond to the data for \textit{BH1} in Figs. \ref{DphiAll} and \ref{TobsAll} classified in three categories. The green and blue lines correspond to the trajectories that asymptotically approach the outer and inner photon spheres, with impact parameters close to $b_{cr_3}$ and $b_{cr_1}$, respectively, while the red line corresponds to trajectories that orbit between the two photon spheres, as illustrated in Fig. \ref{Set1:Between r1 and r3}. From the figure, we infer that trajectories covering the same angular distance $\Delta\phi$ with $n\leq 11\pi$ forming the triplets are observed in the following time-sequence: (i) the trajectory with $b>b_{cr_3}$ is observed first, outside the outer photon ring; (ii) the trajectory with $b_{cr_1}<b\leq b_{min}$ will be observed next, close to the inner photon ring; (iii) the trajectory with $b_{min}<b<b_{cr_3}$ is observed at last between the two photon rings.

\end{enumerate}

In the next section, we examine the trajectories in \textit{BH2} spacetime and find that the time sequence followed by the triplets covering the same angular distance $\Delta\phi$ in \textit{BH2} is different from that in \textit{BH1} spacetime beyond a certain value of $\Delta\phi$.

\subsection{\textit{BH 2}}

We next carry out a similar analysis for \textit{BH2}, as performed for \textit{BH1} in the previous subsection. In particular, we identify the impact parameters of the photon trajectories connecting the source and observer in \textit{BH2} spacetime and compute the quantities introduced at the beginning of section \ref{Time delay of BHs}. The evaluated values for each trajectory are listed in Table \ref{Set2 Data} below.

\begin{table}[h]
\centering
\begin{tabular}{|c|c|c|c|c|c|c|c|c|}
\hline
\multicolumn{8}{|c|}{\textit{BH2}} \\
\hline
No. & $b$ & $r_{turn}$ & $\Delta \phi$ & $T_{obs}$ & $n$ & $ n_{out}$ & $n_{in}$ \\
\hline
 1 & 7.07107 & 1.54128 & 29.0632 & 315.804 & 9 & 0 & 9 \\
 \hline
 2 & 7.07108 & 1.54168 & 27.4914 & 304.69 & 8 & 0 & 8 \\
 \hline
 3 & 7.07193 & 1.5471 & 22.7771 & 271.38 & 7 & 0 & 7 \\
 \hline
 4 & 7.07439 & 1.55325 & 21.2064 & 260.244 & 6 & 0 & 6 \\
 \hline
 5 & 8.08629 & 1.88434 & 21.2057 & 263.138 & 6 & 1 & 5 \\
 \hline
 6 & 8.19233 & 1.9164 & 22.7765 & 275.928 & 7 & 2 & 5 \\
 \hline
 7 & 8.35587 & 1.97062 & 27.4888 & 314.983 & 8 & 1 & 7 \\
 \hline
 8 & 8.38278 & 1.98032 & 29.0596 & 328.131 & 9 & 2 & 7 \\
 \hline
 9 & 8.4262 & 1.99658 & 33.772 & 367.753 & 10 & 3 & 7 \\
 \hline
 10 & 8.43343 & 1.99937 & 35.3428 & 380.994 & 11 & 3 & 8 \\
 \hline
 11 & 8.44498 & 2.00387 & 40.0547 & 420.764 & 12 & 3 & 9 \\
 \hline
 12 & 8.44687 & 2.00461 & 41.6223 & 434.003 & 13 & 4 & 9 \\
 \hline
 13 & 8.45157 & 3.58871 & 29.0612 & 320.81 & 9 & 9 & 0 \\
 \hline
 14 & 8.45159 & 3.59213 & 27.4902 & 307.532 & 8 & 8 & 0 \\
 \hline
 15 & 8.45192 & 3.61339 & 22.7773 & 267.701 & 7 & 7 & 0 \\
 \hline
 16 & 8.4523 & 3.62702 & 21.2064 & 254.423 & 6 & 6 & 0 \\
 \hline
 17 & 8.45754 & 3.71313 & 16.493 & 214.576 & 5 & 5 & 0 \\
 \hline
 18 & 8.46357 & 3.76972 & 14.9225 & 201.289 & 4 & 4 & 0 \\
 \hline
 19 & 8.55475 & 4.16339 & 10.2101 & 161.264 & 3 & 3 & 0 \\
 \hline
 20 & 8.67305 & 4.46749 & 8.63933 & 147.746 & 2 & 2 & 0 \\
 \hline
 21 & 12.0902 & 8.90724 & 3.92775 & 102.673 & 1 & 1 & 0 \\
 \hline
 22 & 19.585 & 16.7565 & 2.35619 & 78.698 & 0 & 0 & 0 \\
 \hline
 \end{tabular}
 \caption{Impact parameters $b$ of photon trajectories connecting the source and observer in \textit{BH2} spacetime along with the turning point $r_{turn}$, angular distance covered $\Delta \phi$, time of observation $T_{obs}$, total number of half-orbits $n$, number of orbits outside the outer photon sphere $n_{out}$ and number of orbits between the two photon spheres $n_{in}$.}\label{Set2 Data}
\end{table}

Since the photon trajectories in \textit{BH2} exhibit the same qualitative structure as those in \textit{BH1}, we do not reproduce the corresponding plots here. Instead, we classify the trajectories in \textit{BH2} according to their structural similarity with those in \textit{BH1}. Trajectories numbered $1-4$ have impact parameters close to $b_{cr_1}$ and orbit only the inner photon sphere, thereby having $n_{out}=0$, and are structurally analogous to the trajectories shown in Fig. \ref{Set1:Close to r1}. Trajectories numbered $5-12$ have impact parameters in the range $b_{cr_1}<b<b_{cr_3}$ and orbit both the inner and outer photon spheres, with nonzero $n_{in}$ and $n_{out}$, exhibiting structures similar to those presented in Fig. \ref{Set1:Between r1 and r3}. Finally, trajectories numbered $13-22$ with $b>b_{cr_3}$ orbit only the outer photon sphere with $n_{in}=0$, and display structures analogous to those shown in Figs. \ref{Set1:Outside r3} and \ref{Set1:Outside r3 magnified}.

For all these trajectories, we next plot the time of observation $T_{obs}$ and the angular distance covered $\Delta\phi$ against the impact parameter $b$ in Figs. \ref{Set2: Tvsb} and \ref{Set2: Phivsb} below.
\begin{figure}[ht]
    \centering
     \begin{subfigure}[c]{0.45\textwidth}
         \centering
         \includegraphics[scale=0.60]{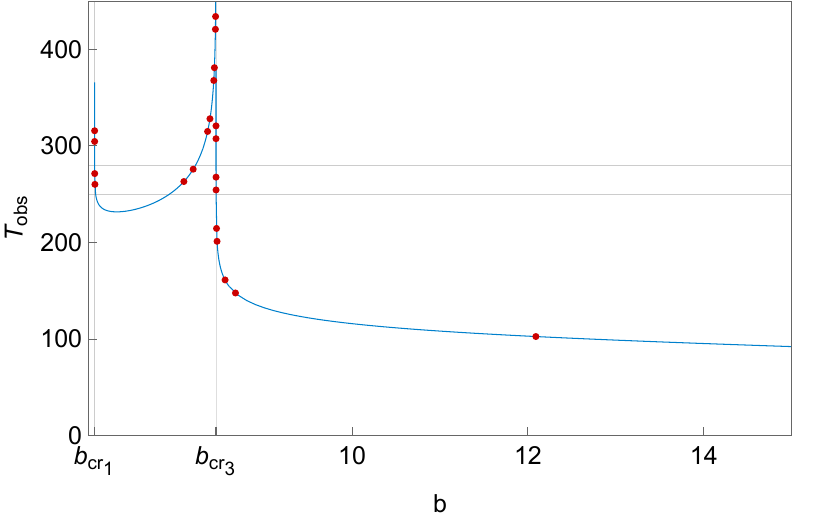}
         \caption{}
         \label{Set2: Tvsb}
     \end{subfigure}
     \hfill
     \begin{subfigure}[c]{0.45\textwidth}
          \centering
          \includegraphics[scale=0.60]{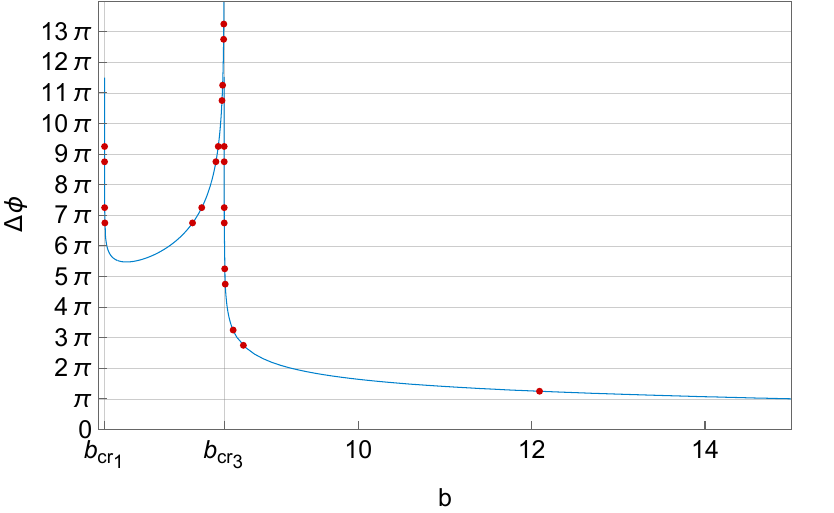}
          \caption{}
          \label{Set2: Phivsb}
     \end{subfigure}
     \caption{The observation time $T_{obs}$ and the angular distance covered $\Delta\phi$ of trajectories connecting source an observer in \textit{BH2}. The red points show the data from Table \ref{Set2 Data} and the blue line shows the data plotted in Figs. \ref{DphiAll} and \ref{TobsAll} for \textit{BH2} where the sources were uniformly distributed on a circle of radius $R_s$. The two horizontal lines in the left figure denote $T_{obs}=250$ and $T_{obs}=280$.}
\end{figure}
The red points denote the data in Table \ref{Set2 Data} while the blue curve corresponds to the data shown in Figs. \ref{DphiAll} and \ref{TobsAll} for \textit{BH2}. As discussed in the case of \textit{BH1}, each pair of the red points represents the images appearing on the opposite sides of the black holes. From the figure, we note that the $6^{th}$ and $7^{th}$ order images associated with the impact parameters $b=8.08629$ and $b=8.19233$, respectively, appear between the two photon rings within the interval $260<T_{obs}<280$. These images therefore reach the observer earlier than the corresponding images in the \textit{BH1} background. This observation is in agreement with the conclusion drawn from Fig. \ref{TobsAll} in Section \ref{comparison with SCh}, that images in a black hole spacetime with a shallower potential well appear at earlier observation times. The $6^{th}$ and $7^{th}$ order images can be distinctly observed between the two photon rings, and the corresponding observation times and impact parameters can provide crucial information about the effective potential and the underlying geometry.

Further, we have plotted the time of observation $T_{obs}$ against the angular distance covered $\Delta\phi$ for the data in Table \ref{Set2 Data}, in Fig. \ref{Set2: TvsPhi} below.
\begin{figure}[h!]
    \centering
    \begin{subfigure}[c]{0.45\textwidth}
         \centering
         \includegraphics[scale=0.55]{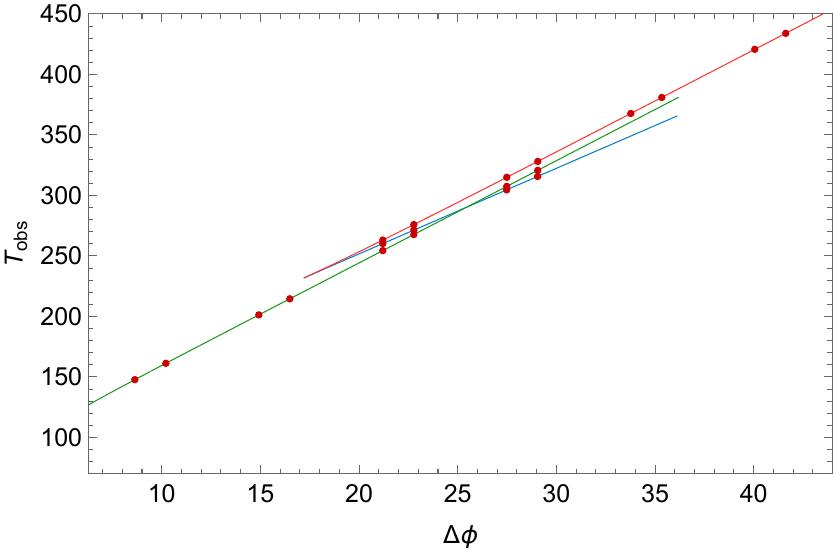}
         \label{Set2: TvsPhi full}
     \end{subfigure}
     \hfill
     \begin{subfigure}[c]{0.45\textwidth}
          \centering
          \includegraphics[scale=0.55]{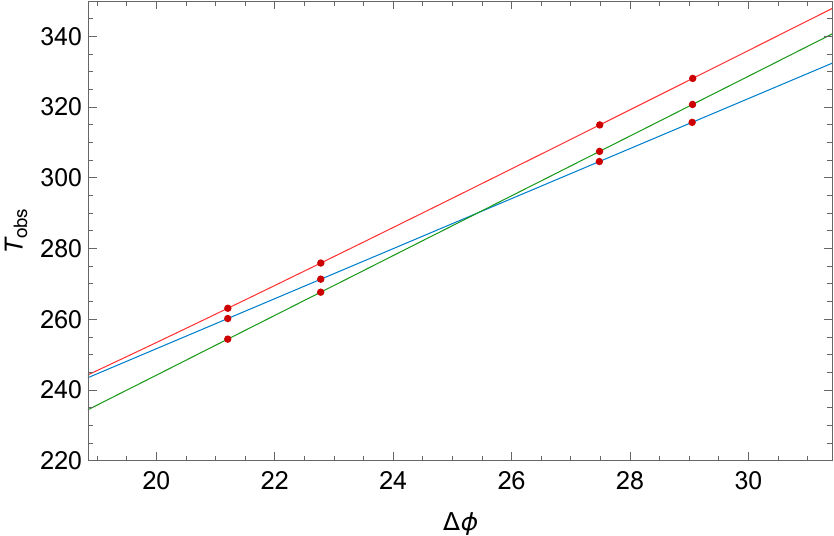}
          \label{Set2: TvsPhi magnified}
     \end{subfigure}
    \caption{The red points represent the data in Table \ref{Set2 Data}. The three lines correspond to the data for \textit{BH2} in Figs. \ref{DphiAll} and \ref{TobsAll} classified as: (i) Green line: Trajectories outside the outer photon sphere with $b>b_{cr_3}$, (ii) Blue line: Trajectories close to the inner photon sphere with $b_{cr_1}<b\leq b_{min}$, (iii) Red line: Trajectories falling between the two photon spheres with $b_{min}<b<b_{cr_3}$. The figure on the right shows the enlarged view of the intersection of the blue and green lines.}
    \label{Set2: TvsPhi}
\end{figure}
Similar to Fig. \ref{Set1: TvsPhi}, the red points denote the data listed in Table \ref{Set2 Data}, while the green, blue, and red curves correspond to the data shown in Figs. \ref{DphiAll} and \ref{TobsAll}. These curves represent trajectories with $b>b_{cr_3}$, $b_{cr_1}<b\leq b_{min}$ and $b_{min}<b<b_{cr_3}$ respectively.

We highlight here that only the $6^{th}$ and $7^{th}$ order trajectories follow the time-sequence of observations discussed at the end of the previous section in point number 8 in the case of \textit{BH1}. The green and blue curves intersect near $\Delta\phi\approx 25$; beyond which the trajectories contributing to the outer photon ring arrive at the observer's location later than those associated with the inner photon ring. Notably, this reversal in the time-sequence of images occurs at an earlier observation time and at a smaller $\Delta\phi$ in the \textit{BH2} spacetime than in the case of \textit{BH1}. This behavior indicates that the higher-order images beyond a certain order appear first in the vicinity of the inner photon ring and subsequently near the outer photon ring. Thus, the image order at which this transition occurs can serve as a probe of the depth of the potential well and can resolve the degeneracy between the parameter space, such as in \textit{BH1} and \textit{BH2}. For the range of $\Delta\phi$ considered in Fig. \ref{Set2: TvsPhi}, the trajectories with $b_{min}<b<b_{cr_3}$ will still be observed at last between the two photon rings, similar to the case of \textit{BH1}.

The investigations so far of strongly lensed image positions and associated time delays have been carried out using the metric defined in Eq. (\ref{Metric}), with the metric function given in Eq. (\ref{metric function}). However, as discussed in Section \ref{double peak potentials}, the effective potentials shown in Fig. \ref{Potentials for 2 sets} differ in the magnitudes of the potential at the anti-photon spheres, as well as in the locations of the photon and anti-photon spheres. Consequently, the observed image positions and time delays are influenced by both the smaller/larger turning points and the deeper/shallower potential wells. In order to isolate the effect of the depth of the potential well, in the next section, we consider a spacetime described by a modified metric function, constructed such that, apart from the depth of the potential well, all other relevant quantities, namely the locations of the photon and anti-photon spheres and the magnitudes of the potentials at the peaks, are held fixed.

\section{Exotic Compact Object}\label{ECO}

In this section, we choose a metric function and the parameter values such that the resulting potential has extrema, that is, the two peaks and the minimum, at the same location as that for \textit{BH1}, thus fixing the locations of photon and anti-photon spheres. Further, the choice ensures that the magnitude of potential at the peaks is the same as in the case of \textit{BH1}. Crucially, the only difference in the potential outside the inner photon sphere is the magnitude of the potential at the location of the anti-photon sphere. 

Such a construction allows one to explicitly analyze the effect of the depth of the potential minimum. We note here that the potential constructed in this section asymptotes to a large value inside the inner photon sphere, which is the behavior associated with a spacetime of an exotic compact object (ECO). We thus refer to the potential as ECO potential. Since we are exploring the photon trajectories outside the inner photon sphere, with $b>b_{cr_1}$, the imprints of potential well depths are not impacted by the internal structure of the potential. Hence, the ECO potential is suitable for studying and isolating the effects of different potential well depths on the photon trajectories.

We choose the spacetime metric defined by Eq.(\ref{Metric}) with the metric function given by
\begin{eqnarray}
    f(r) &=& 1-\frac{A}{r}+\frac{B}{r^2}-\frac{C}{r^3}+\frac{D}{r^4}-\frac{E}{r^5}+\frac{F}{r^6} \label{ECO metric function}
\end{eqnarray}
where $A,B,C,D,E,F$ are constant parameters defining the ECO spacetime. The choice of $A=5.45478$, $B=15.6107$, $C=38.9757$, $D=72.6156$, $E=71.6123$, and $F=27.0054$ then gives a potential that differs from the potential of \textit{BH1} only in terms of the magnitude of potential at the anti-photon sphere in our region of interest, i.e., outside the inner photon sphere. The potential in ECO spacetime, along with the potential of \textit{BH1}, is illustrated in Fig. \ref{ECO potential} below.
\begin{figure}[h!]
    \centering
    \includegraphics[width=0.6\linewidth]{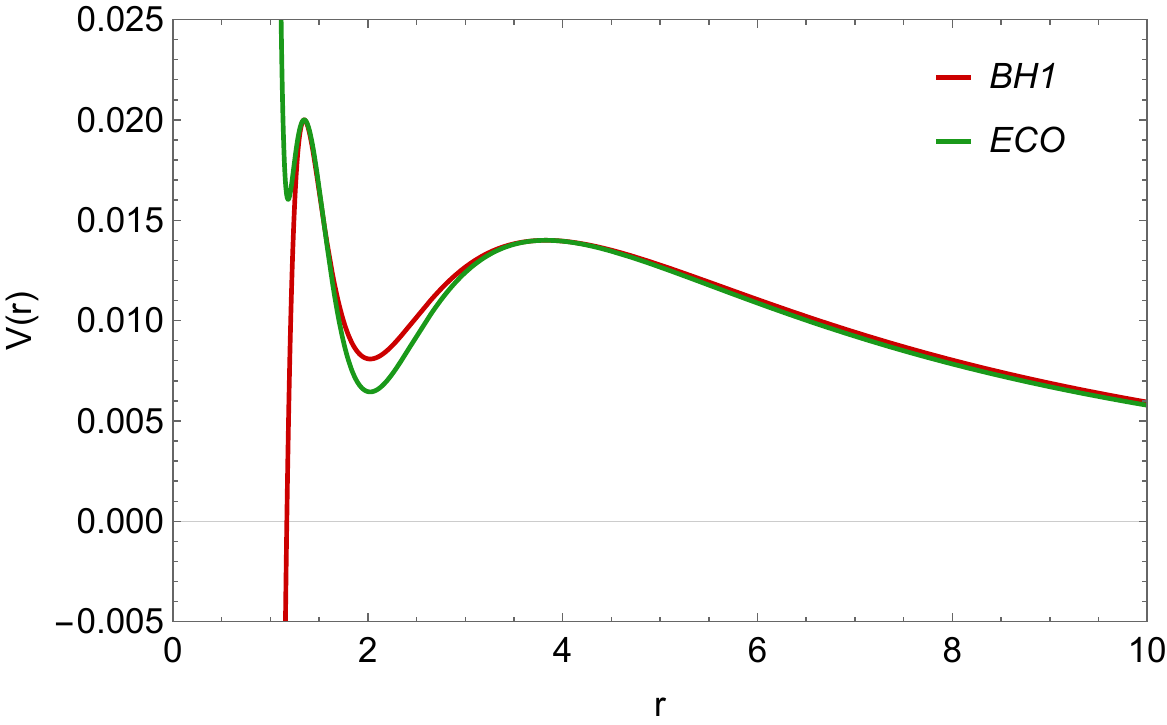}
    \caption{The green and red curves show the ECO and \textit{BH1} potentials respectively.}
    \label{ECO potential}
\end{figure}
We now follow the same steps of analysis for the ECO potential as in the \textit{BH1} case discussed in the previous sections. Considering the point sources to be located uniformly on a circle of radius $R_s=15\sqrt{2}$ around the black hole and the observer at $r_{obs}=50$, we evaluated the total angular distance $\Delta\phi$ covered and time taken $T_{obs}$ by the photon trajectories to reach a circle of radius $R_s$ starting from the observer's location by backward ray tracing, as obtained for \textit{BH1} spacetime in Section 4. The data obtained are presented in Figs. \ref{ECO:DphiAll} and \ref{ECO:TobsAll} below, along with the corresponding data of \textit{BH1}.
\begin{figure}[h!]
    \centering
    \begin{subfigure}[c]{0.45\textwidth}
          \centering
          \includegraphics[scale=0.55]{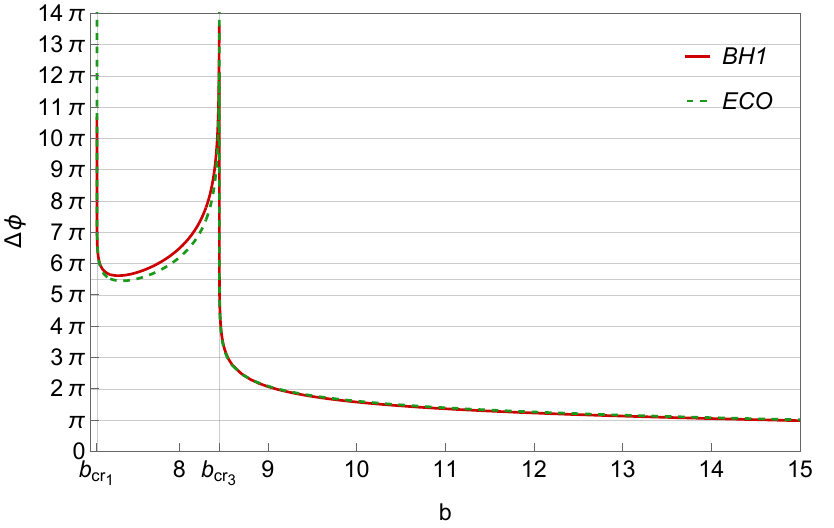}
          \caption{}
          \label{ECO:DphiAll}
     \end{subfigure}
     \hfill
     \begin{subfigure}[c]{0.45\textwidth}
         \centering
         \includegraphics[scale=0.55]{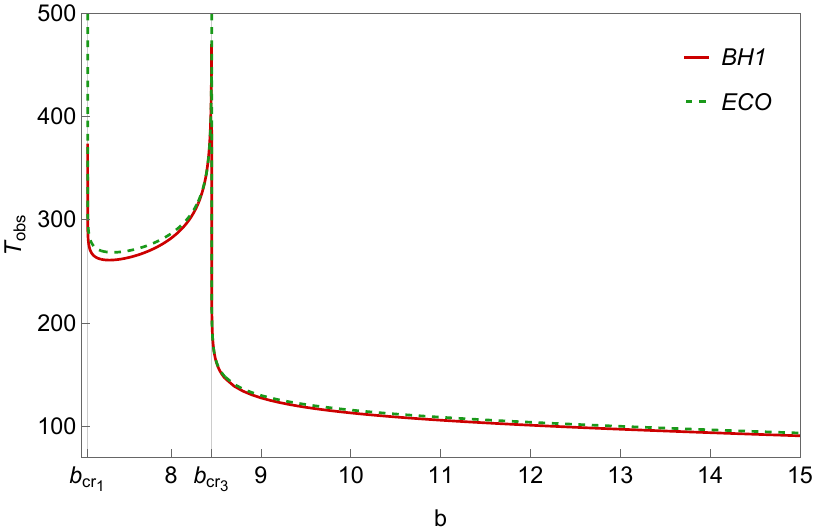}
         \caption{}
         \label{ECO:TobsAll}
     \end{subfigure}
     \caption{The angular distance covered $\Delta\phi$ and the observation time $T_{obs}$ for all trajectories connecting the sources uniformly distributed in a circle of radius $R_s=15\sqrt{2}$ and observer at $r_{obs}=50$ in the background of ECO and \textit{BH1}.}
\end{figure}
The minimum angular distance covered, $\phi_{\min}$, in the ECO case is smaller than that in \textit{BH1}, while the time taken by the corresponding trajectory is larger in the ECO background. This suggests that, for trajectories with $b_{cr_1}<b<b_{cr_2}$, a deeper potential well leads to a smaller coverage of angular distance accompanied by a longer travel time.

\begin{figure}[h!]
    \centering
     \begin{subfigure}[c]{0.45\textwidth}
         \centering
         \includegraphics[scale=0.65]{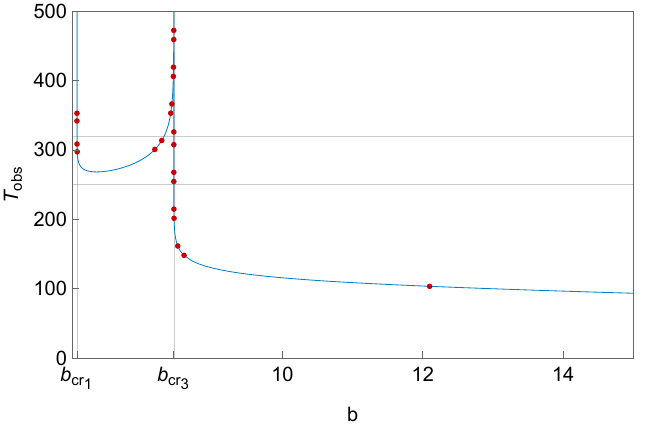}
         \caption{}
         \label{ECO: Tvsb}
     \end{subfigure}
     \hfill
     \begin{subfigure}[c]{0.45\textwidth}
          \centering
          \includegraphics[scale=0.65]{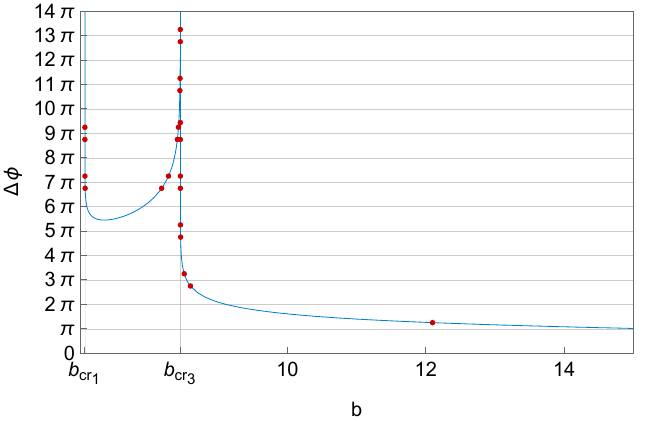}
          \caption{}
          \label{ECO: Phivsb}
     \end{subfigure}
     \caption{Total angular distance covered $\Delta\phi$ and time of observation $T_{obs}$ for the multiple images of the source observed at different impact parameters. The data from Table \ref{ECO data} is plotted as red points. The blue line shows the data plotted in Figs. \ref{ECO:DphiAll} and \ref{ECO:TobsAll} for ECO. The two horizontal lines in the left figure denote $T_{obs}=250$ and $T_{obs}=320$.}
\end{figure}

We next find the trajectories connecting the source and observer located at $(R_s,\phi_s)=(15\sqrt{2},\,3\pi/4)$ and $(r_{obs},\phi_{obs})=(50,\,0)$ respectively, and further evaluate the corresponding quantities as defined in section \ref{Time delay of BHs} for each trajectory. The values thus obtained are listed in Table \ref{ECO data} below. 
\begin{table}[h!]
\centering
\begin{tabular}{|c|c|c|c|c|c|c|c|c|}
\hline
\multicolumn{8}{|c|}{ECO} \\
\hline
No. & $b$ & $r_{turn}$ & $\Delta \phi$ & $T_{obs}$ & $n$ & $ n_{out} $& $n_{in}$ \\
\hline
 1 & 7.07107 & 1.34783 & 29.0442 & 352.862 & 9 & 0 & 9 \\
 \hline
 2 &7.07109 & 1.3482 & 27.4777 & 341.784 & 8 & 0 & 8 \\
 \hline
 3 & 7.07209 & 1.35276 & 22.7724 & 308.512 & 7 & 0 & 7 \\
 \hline
 4 & 7.07483 & 1.3577 & 21.2027 & 297.409 & 6 & 0 & 6 \\
 \hline
 5 & 8.1808 & 1.54739 & 21.205 & 300.699 & 6 & 1 & 5 \\
 \hline
 6 & 8.27957 & 1.5573 & 22.7758 & 313.633 & 7 & 2 & 5 \\
 \hline
 7 & 8.40773 & 1.56975 & 27.4882 & 353.018 & 8 & 1 & 7 \\
 \hline
 8 & 8.42391 & 1.57129 & 29.0593 & 366.287 & 9 & 2 & 7 \\
 \hline
 9 & 8.4446 & 1.57325 & 33.7713 & 405.994 & 10 & 3 & 7 \\
 \hline
 10 & 8.44717 & 1.57349 & 35.3421 & 419.261 & 11 & 4 & 7 \\
 \hline
 11 & 8.45045 & 1.5738 & 40.0545 & 459.077 & 12 & 3 & 9 \\
 \hline
 12 & 8.45086 & 1.57384 & 41.6253 & 472.352 & 13 & 4 & 9 \\
 \hline
 13 & 8.45154 & 3.82877 & 29.6575 & 325.992 & 9 & 9 & 0 \\
 \hline
 14 & 8.45154 & 3.82975 & 27.479 & 307.58 & 8 & 8 & 0 \\
 \hline
 15 & 8.45158 & 3.83595 & 22.7715 & 267.794 & 7 & 7 & 0 \\
 \hline
 16 & 8.45163 & 3.84082 & 21.202 & 254.529 & 6 & 6 & 0 \\
 \hline
 17 & 8.45289 & 3.88089 & 16.4913 & 214.715 & 5 & 5 & 0 \\
 \hline
 18 & 8.45495 & 3.91274 & 14.9219 & 201.491 & 4 & 4 & 0 \\
 \hline
 19 & 8.50758 & 4.19372 & 10.2099 & 161.587 & 3 & 3 & 0 \\
 \hline
 20 & 8.5972 & 4.44706 & 8.63911 & 148.157 & 2 & 2 & 0 \\
 \hline
 21 & 12.0972 & 8.87554 & 3.92684 & 103.481 & 1 & 1 & 0 \\
 \hline
 22 & 19.815 & 16.8616 & 2.35615 & 79.2952 & 0 & 0 & 0 \\
\hline
\end{tabular}
\caption{Impact parameters $b$ of photon trajectories connecting the source and observer in the ECO spacetime along with the turning point $r_{turn}$, time of observation $T_{obs}$, angular distance covered $\Delta \phi$, total number of half-orbits $n$, number of orbits outside the outer photon sphere $n_{out}$ and number of orbits between the two photon spheres $n_{in}$.}
\label{ECO data}
\end{table}
These trajectories have features similar to those in \textit{BH1} and \textit{BH2}; hence, we do not construct plots of the entire trajectories here. However, we plotted the angular distance covered $\Delta\phi$ and the time of observation $T_{obs}$ against the impact parameter $b$ in Figs. \ref{ECO: Phivsb} and \ref{ECO: Tvsb} below for all the trajectories in Table \ref{ECO data}.
Similar to the case of \textit{BH1}, the $6^{th}$ and $7^{th}$ order images appear within the time interval $250<T_{obs}<320$. Thus, within this time window, one can observe the triplets of $n=6$ and $n=7$.

Further, to understand how the observation time $T_{obs}$ varies with the angular distance covered $\Delta\phi$, we have plotted the relevant data in Fig. \ref{comparison of ECO and BH1} below.
\begin{figure}[h!]
    \centering
    \includegraphics[scale=0.70]{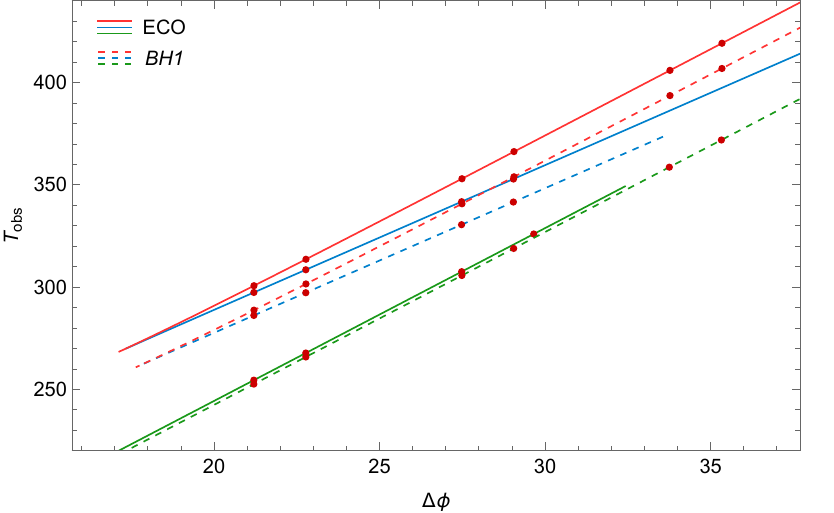}
    \caption{The time of observation $T_{obs}$ against the angular distance covered $\Delta\phi$ for ECO and \textit{BH1}. The solid lines show ECO data while the dashed lines show the data of \textit{BH1} along with the red points representing the data in Tables \ref{Set1 Data} and \ref{ECO data}.}
    \label{comparison of ECO and BH1}
\end{figure}
The three lines green, blue, and red correspond to the three categories of trajectories with $b>b_{cr_3}$, $b_{cr_1}<b\leq b_{min}$, and $b_{min}<b<b_{cr_3}$, respectively, as defined in the previous section. From the figure, it is evident that the images follow the same time-sequence as in the case of \textit{BH1}. However, the trajectories covering the same angular distance $\Delta\phi$ take a longer time in the ECO background compared to \textit{BH1}, irrespective of their category. This emphasizes the conclusion that spacetime with a shallower potential well at the anti-photon sphere leads to a shorter travel time for photon trajectories between the two photon spheres.

\section{Discussion}\label{discussion section}


We have shown that the time delays associated with the higher-order images of a transient source around a black hole provide a promising observable for resolving parameter degeneracies in spacetimes admitting multiple photon spheres. We investigate photon trajectories within a parametrized, static, and spherically symmetric theory-agnostic framework. We found that these trajectories are characterized by the nontrivial temporal behavior, including a minimum travel time, a minimum angular deflection, and a characteristic triplet structure of higher-order images with a specific arrival sequence. Since the geodesic motion of a photon is primarily affected by the potential encountered by it, these characteristic features of time delay will be broadly applicable to any spacetime metric that exhibits a similar underlying potential structure. In this sense, our conclusions are model-independent. Our approach, therefore, allows us to identify generic time-delay features common to spacetimes with double-peak potential profiles. The techniques used in this work can be applied to any spacetime solution to obtain time delay estimates relevant for testing black hole environments or gravity theories.

To investigate parameter degeneracies, we analyzed two distinct spacetime configurations that produce identical photon ring radii in the black hole images. We then comprehensively examined the photon geodesics and quantified them by the total angular distance covered $\Delta\phi$ and the time taken by the trajectories $T_{obs}$, along with the turning point, and the total number $n$ of half($\pi$)-turns. For the photon trajectories falling inside the outer photon sphere, the total number of turns was divided into two parts, namely the turns completed outside the outer photon sphere $n_{out}$, and the turns completed between the two photon spheres $n_{in}$.

Based on the comparison of these quantities with the Schwarzschild case, we arrive at the following observations. In the case of double peak potential, $\Delta\phi$ and $T_{obs}$ decrease monotonically with the impact parameter for trajectories outside the outer photon sphere, similar to the case of the Schwarzschild black hole. However, between the two photon spheres, these quantities initially decrease, starting from the inner photon sphere, to minimum values, say $\phi_{min}$ and $T_{min}$ at impact parameter $b_{min}$, and subsequently increase, approaching large values close to the outer photon sphere. In other words, the photon trajectories outside the outer photon sphere can complete any $n$ number of half turns starting from $n=0$ around the black hole, whereas the trajectories falling between the two photon spheres complete at least a nonzero minimum number of turns, say $N_{min}$ (associated with the angular distance $\phi_{min}$), around the black hole before reaching the observer. The minimum values $\phi_{min}$ and $T_{min}$ correspond to the same trajectory with impact parameter $b_{min}$.

 We further observed a correlation between the depth of the potential well at the anti-photon sphere and the resulting temporal signatures. Among the black hole geometries considered, the spacetime with a shallower potential well at the anti-photon sphere leads to a shorter travel time for photon trajectories between the two photon spheres. In these spacetimes, the effective potentials differ both in the potential values at the anti-photon sphere and in the locations of the photon and anti-photon spheres. Consequently, the observed image positions and time delays are influenced by both the smaller/larger turning points and the deeper/shallower potential wells encountered by the trajectories.
 
Furthermore, for a fixed point source-observer configuration, we identified the connecting trajectories and found that beyond a certain order $n$, there exists a triplet of trajectories for each order $n$, which we categorized as: (i) Trajectories orbiting only the inner photon sphere with $n_{out}=0$. These trajectories contribute to the sub-ring structure of the inner photon ring. (ii) Trajectories orbiting only the outer photon sphere with $n_{in}=0$. These trajectories contribute to the sub-ring structure of the outer photon ring. (iii) Trajectories traversing between the two photon spheres with both nonzero $n_{in}$ and $n_{out}$. The images formed by these trajectories appear between the two photon rings. The division of the total number of turns $n$ among $n_{out}$ and $n_{in}$ is expected to depend on the relative positions of the source and observer.

We find that the images associated with these trajectory triplets reach the observer within a narrow temporal window and follow a well-defined time sequence up to a certain order, say $n = N$. Consequently, an appropriately chosen time window can capture multiple images of the source at distinct impact parameters. For each order $n$, the image appears first close to the outer photon sphere (corresponding to $n_{in}=0$), then close to the inner photon sphere (corresponding to $n_{out}=0$), and at last in between the two photon spheres (corresponding to $n_{in}\neq 0$ and $n_{out}\neq 0$). Beyond $n=N$, the sequence of the first two images gets reversed, that is, the image appears first in the vicinity of the inner photon ring and subsequently near the outer photon ring. The image order at which this transition occurs can serve as an additional probe of the background spacetime.

We further analyzed the trajectories in the background of an exotic compact object where the potential, up to the inner peak, differs from that of the black hole spacetime only in terms of the depth of the potential well at the anti-photon sphere while all other relevant quantities, namely the locations of the photon and anti-photon spheres and the magnitudes of the potentials at the peaks, are same as in the black hole case. This allowed us to analyze the isolated effect of the different depths of the potential well. We found that for trajectories traversing between the two photon spheres, a deeper potential well leads to a smaller coverage of angular distance, accompanied by a longer travel time.

Overall, our results emphasize that time delay observables probe the full structure of the effective potential and can reveal features that remain hidden in static shadow images. In particular, they offer direct access to the region between the photon spheres, providing a new handle on the underlying spacetime geometry. We note, however, that current very long baseline interferometric observations, such as those by the Event Horizon Telescope, reconstruct images from data integrated over finite time intervals, effectively averaging over rapid variability. As a result, the direct temporal ordering of individual lensed images may not be readily accessible in standard observations. Nevertheless, with the development of time-dependent imaging techniques and improved temporal resolution in next-generation instruments, these signatures may become observable, either through direct time-resolved measurements or through correlated variability in the observed emission, thereby opening a promising avenue for testing the nature of compact objects and the strong-field regime of gravity.

\section*{Acknowledgment}

This research was initiated at the International Centre for Theoretical Sciences (ICTS) during participation in the program Beyond the Horizon: Testing the black hole paradigm (code: ICTS/beyondhorizon2025/03). We thank Sumanta Chakraborty for insightful discussions, which significantly contributed to the development of the ideas underlying this work. We thank the anonymous referee for useful comments. S.K.’s research is supported by the Anusandhan National Research Foundation, India, under the ANRF ARG Grant (No. ANRF/ARG/2025/005821/PS).

\end{document}